\documentclass{article}
\usepackage{graphicx}
\usepackage{color}

\usepackage{amsmath}
\usepackage{amsfonts}
\usepackage{amssymb}
\usepackage{amsthm}
\usepackage{setspace}

\newtheorem{theorem}{Theorem}

\def\fw{Furedi-Wetzel}
\def\sw{Schaer-Wetzel}

\def\emph#1{\textbf{\textit{#1}}}

\def\figs#1#2#3{%
 \begin{figure}[tp]%
  \centerline{\includegraphics[scale=#3]{#1.eps}}%
  \caption{#2}%
  \label{fig:#1}%
 \end{figure}
 }
\def\figref#1{Figure \ref{fig:#1}}

\begin{document}

\title{A smaller cover for closed unit curves}
\author{Wacharin Wichiramala}
\date{\today}
\maketitle

\begin{abstract}
Forty years ago Schaer and Wetzel showed that a $\frac{1}{\pi}\times\frac
{1}{2\pi}\sqrt{\pi^{2}-4}$ rectangle, whose area is about $0.122\,74,$ is the
smallest rectangle that is a cover for the family of all closed unit arcs.
\ More recently F\"{u}redi and Wetzel showed that one corner of this rectangle
can be clipped to form a pentagonal cover having area $0.11224$ for this
family of curves. \ Here we show that then the opposite corner can be clipped
to form a hexagonal cover of area less than $0.11023$ for this same family.
\ This irregular hexagon is the smallest cover currently known for this family
of arcs.
\end{abstract}

\noindent keywords: covering of unit arcs, covering by convex sets, worm
problem



\section{Introduction}

\begin{figure}[htb] \centering
\begin{tabular}
[c]{ccc}%
{\includegraphics[
natheight=1.895900in,
natwidth=2.108500in,
height=1.1281in,
width=1.251in
]%
{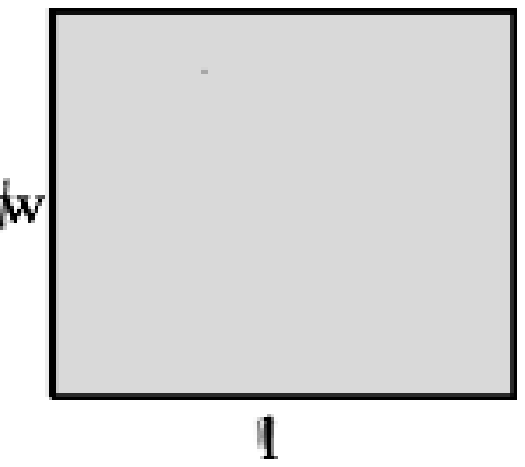}%
}
& $\quad$ &
{\includegraphics[
natheight=3.940100in,
natwidth=4.362100in,
height=1.1283in,
width=1.2453in
]%
{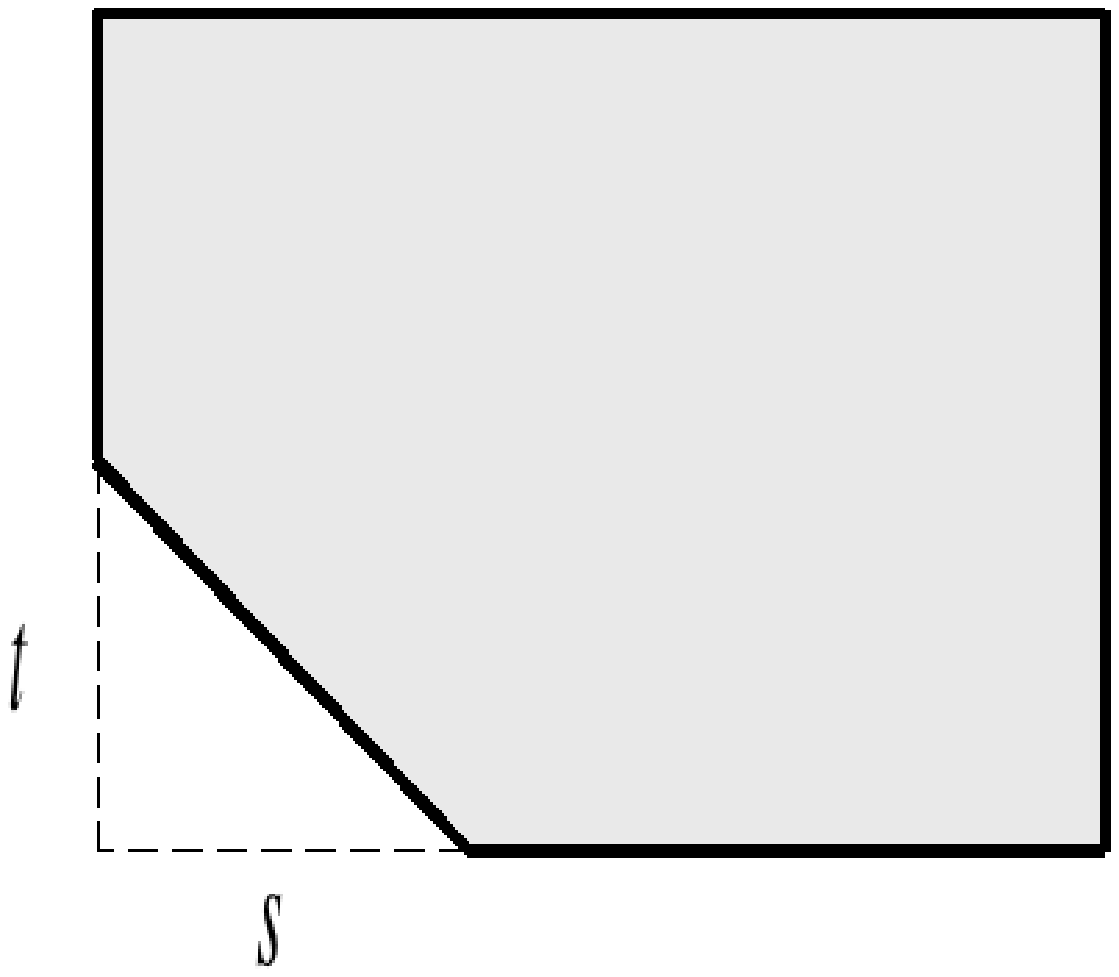}%
}
\\
a. \ Cover $\mathcal{R}$. &  & b. \ Clipped $\mathcal{R}$.\\
Area
$<$
$0.12274.$ &  & Area
$<$%
\ $0.11222.$%
\end{tabular}
\caption{Two covers for $\mathcal{F}_{0}$.}\label{first}%
\end{figure}

Forty years ago Schaer and Wetzel \cite{Schaer-Wetzel} (and independently
Chakarian and Klamkin \cite{Chakerian-Klamkin}) showed that the smallest
rectangular region $\mathcal{R}$ that is a cover for the family $\mathcal{F}%
_{0}$ of all closed unit arcs in the Euclidean plane, that is to say, contains
a congruent (i.e., an isometric) copy of each closed unit arc, is $l=\frac
{1}{\pi}$ by $w=\sqrt{\frac{1}{4}-\frac{1}{\pi^{2}}}$ (Figure \ref{first}%
a.)\ \ This rectangle has area about $0.12274.$

The underlying problem here is a variant for closed unit arcs of a well-known
unsolved problem posed in 1966 by Leo Moser (see \cite[pp. 211, 218-219]%
{Moser}): \ find the area of the smallest (convex) region in the plane that
contains a congruent (i.e., isometric) copy of every arc of unit length. \ The
existence of a convex cover of least area $\alpha_{2}$ for the family
$\mathcal{F}_{0}$ of all closed unit arcs follows from standard compactness
arguments, but its uniqueness is not known. \ Problems of these kinds are
closely related to the well-known Lebesgue Universal Cover problem (see
\cite[\S D15]{Croft-Falconer-Guy}), and they seem equally intractable.

Both \cite{Schaer-Wetzel} and \cite{Chakerian-Klamkin} made the elementary
observation that the least area $\alpha_{2}$ must be greater than the smallest
convex hull of a circle of unit circumference and a unit line segment, a set
having area $0.09632.$ \ Using geometric methods F\"{u}redi and
Wetzel\ \cite{Furedi-Wetzel} recently raised the lower bound to $0.09670, $
and they showed that the pentagonal region formed by suitably clipping one
corner of $\mathcal{R}$ (Figure \ref{first}b) is a cover for $\mathcal{F}_{0}
$ with area less than $0.11222.$%
\ They achieved a further slight reduction in the upper bound by replacing a
portion of the clipping line segment by a short elliptic arc, forming a
curvilinear convex hexagonal cover having area less than $0.11213.$ \ These
are the best published bounds for $\alpha_{2}$ at present:%
\[
0.09670<\alpha_{2}<0.11213.
\]

\section{A smaller cover}

In recent years computational methods have increasingly been employed to
attack geometric questions. \ For example, Brass and his student Sharifi
\cite{Brass-Sharifi} used a grid search numerical method to improve the known
lower bound for the Lebesgue Universal Cover problem. \ In this article we
employ numerical convex optimization to reduce the known upper bound for
$\alpha_{2}$ by nearly $1.7$ per cent. \ More precisely, we establish the
following theorem.

\begin{figure}[htb] \centering
\begin{tabular}
[c]{c}%
{\includegraphics[
natheight=4.722500in,
natwidth=4.351800in,
height=1.5332in,
width=1.4145in
]%
{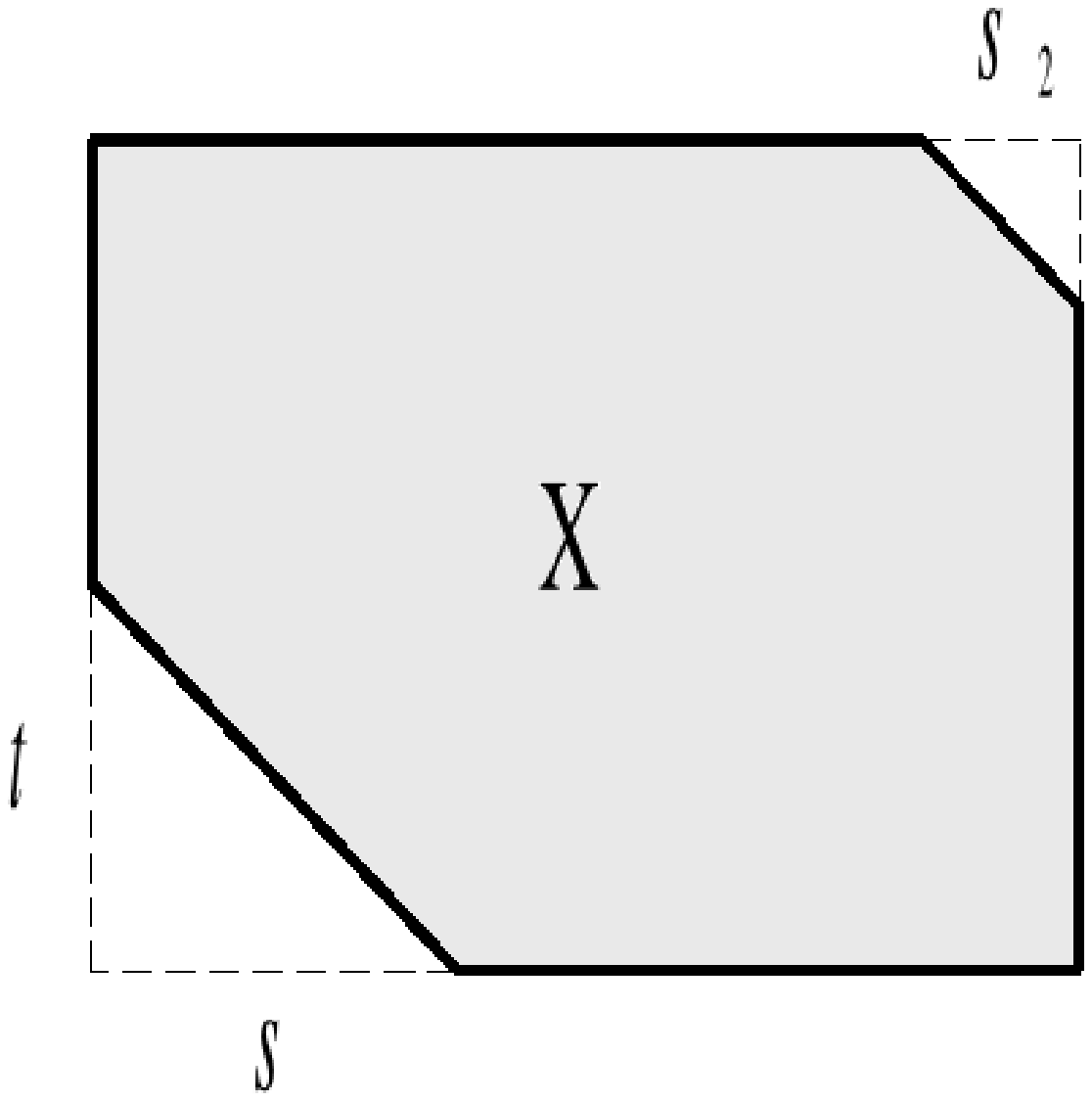}%
}
\\
The cover $\mathfrak{X}.$\\
Area
$<$
$0.11023.$%
\end{tabular}
\caption{A smaller cover for $\mathcal{F}_{0}$.}\label{swbest}%
\end{figure}

\begin{theorem}
Let $s=0.142\,0171,$ $t=0.148\,1552,$ and $s_{2}=0.061\,7141,$ and using these
data, let $\mathfrak{X}$ be the rectangular region $\mathcal{R}$ clipped by
two parallel line segments as pictured in Figure \ref{swbest}. \ Then
$\mathfrak{X}$ is a cover for the family $\mathcal{F}_{0}$ of all closed
curves of unit length, and its area is about $0.110\,2299.$
\end{theorem}%

The area of $\mathfrak{X}$ is $lw-\frac{1}{2}st\left(  1+(s_{2}/s)^{2}\right)
\approx0.110\,2299.$ \ Showing that $\mathfrak{X}$ is a cover for
$\mathcal{F}_{0}$ is the objective of this article.
The value of $s_2$ is chosen so that every closed arc that cannot be covered by $\mathfrak{X}$ must be longer than 1.00001 numerically.

\section{Placing a closed arc in the rectangle}

From \cite{ww}, it suffices for $\mathfrak{X}$ to cover all simple closed unit arcs.
From \cite{d}, every closed unit arc is contained in the convex hull of another
convex closed unit arc. Since $\mathfrak{X}$ is convex, we may assume that
we are dealing with convex closed unit arcs. Since every convex closed unit
arc has diameter at least $1/\pi$, we may place such arc in the rectangle
$\mathcal R$ so that it touches the upper and lower
boundaries \cite{\sw}

\figs{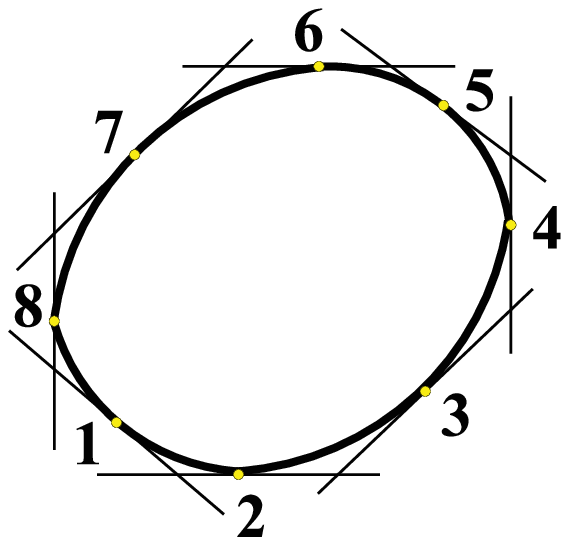}{All touched points $p_1,p_2,...,p_8$.}{0.5}

For convenient, let $\theta=\arctan m$ where $m=t/s$.
Let $p_1,p_2,...,p_8$ be the (chosen) lower left, bottom, lower right, right,
upper right, top, upper left, left points touched by support lines (coming)
from angles $-90^o-\theta,-90^o,...,180^o$ (with slopes $-m,0,m,\infty,-m,0,m,\infty$).
The points $p_i$ appear in counter clockwise order since the closed arc is simple. Note that 2 consecutive points, including $p_1$ and $p_8$, may coincide
(see \figref{all-points}).
For $i=1,...,8$, let $(x_i,y_i)$ be the coordinate of $p_i$.
Let $x_L$ and $x_R$ be the left and right x-coordinate of a concerned rectangle,
called \emph{box}.
As we move the box to the right, we end up with $x_L=x_8$. For the opposite
direction, we have $x_R=x_4$.

When the box is moving to the right (the arc is relatively moving to the
left), the situation that $p_5$ is in the big
corner is called OUTL5
\figs{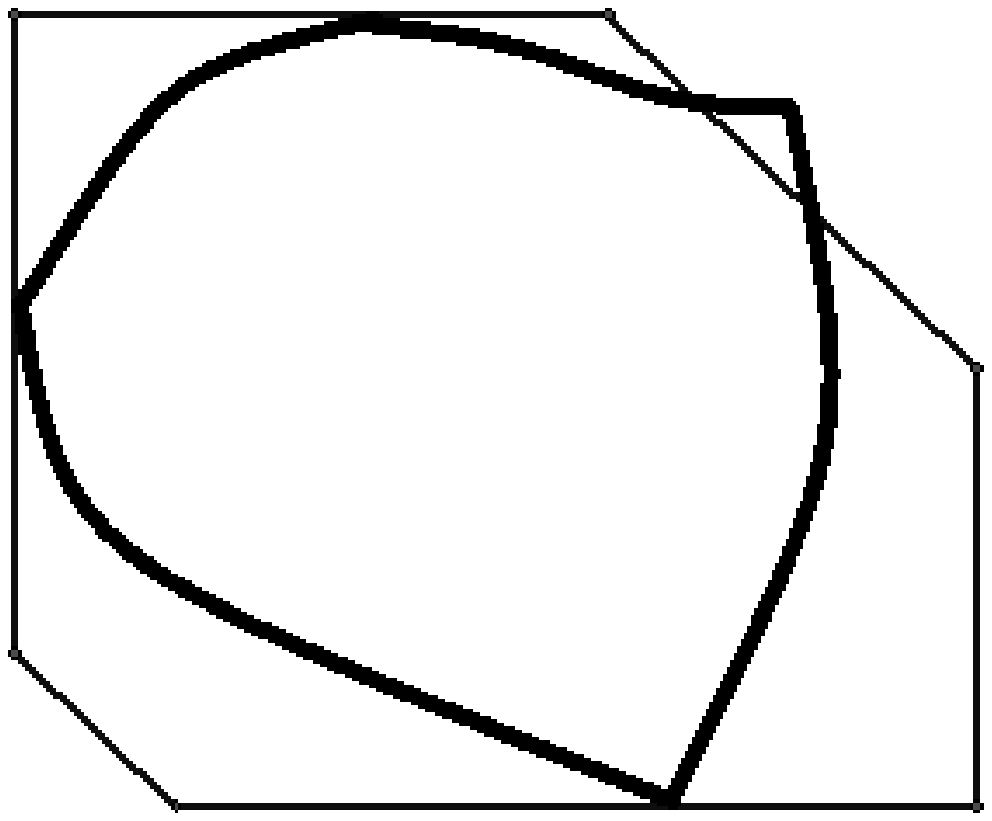}{The situation OUTL5.}{.2}
\figs{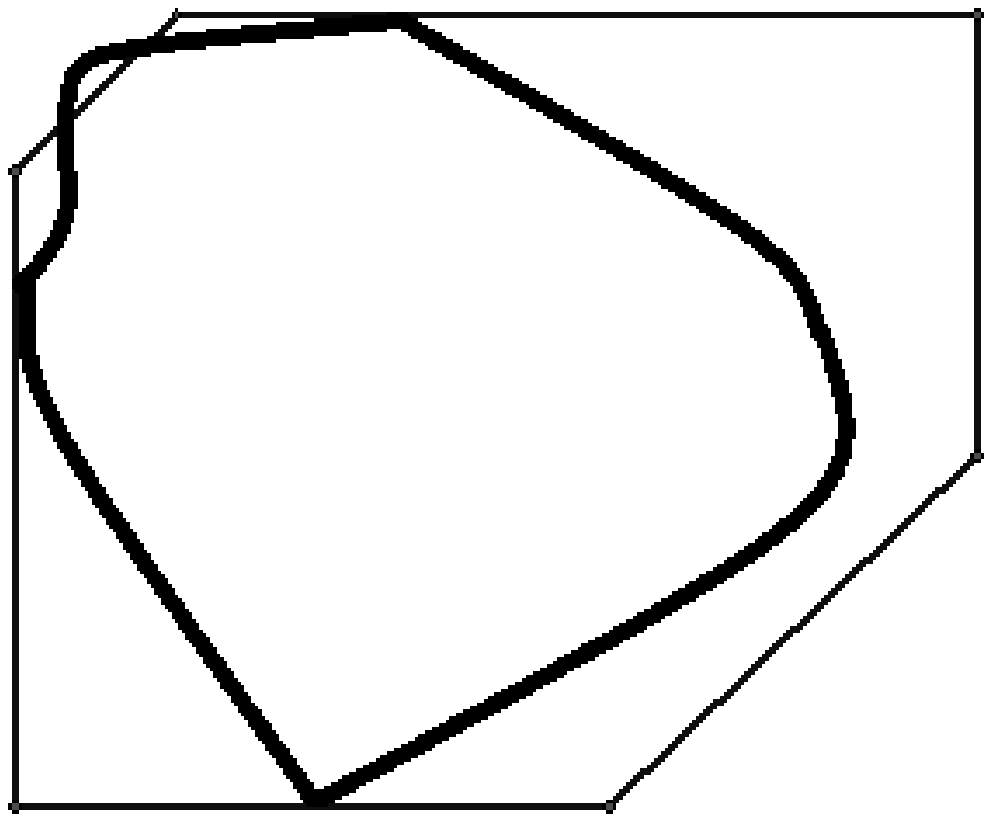}{The situation outL7.}{.2}
\figs{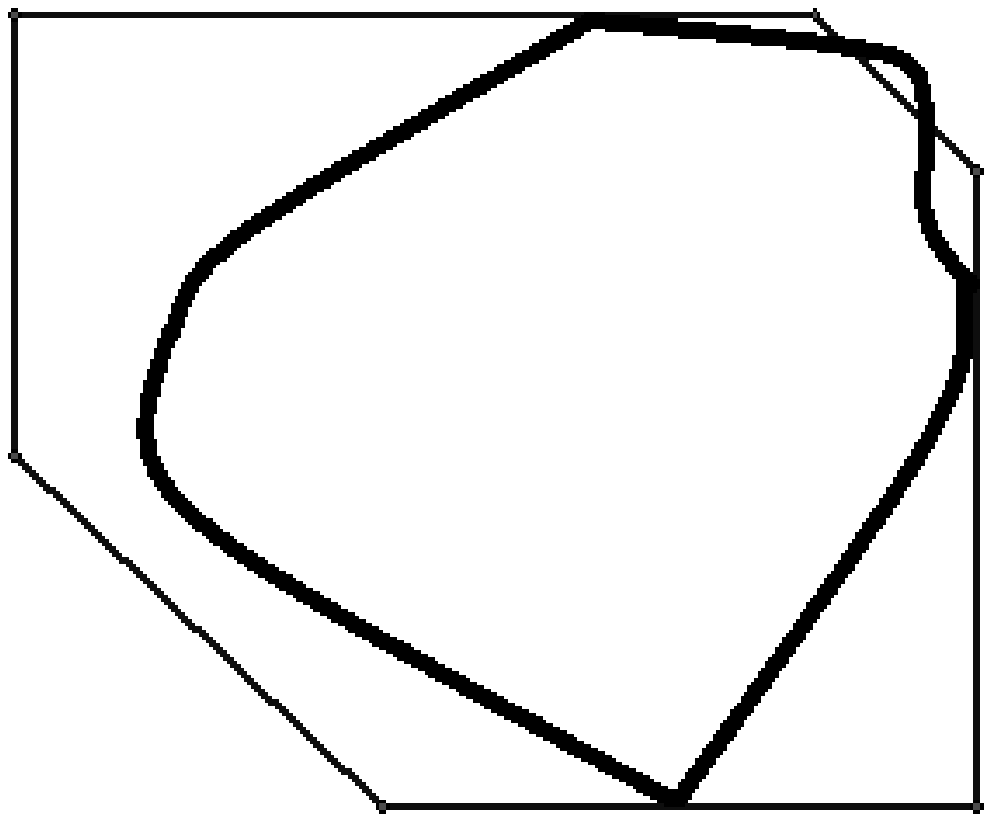}{The situation outR5.}{.2}
\figs{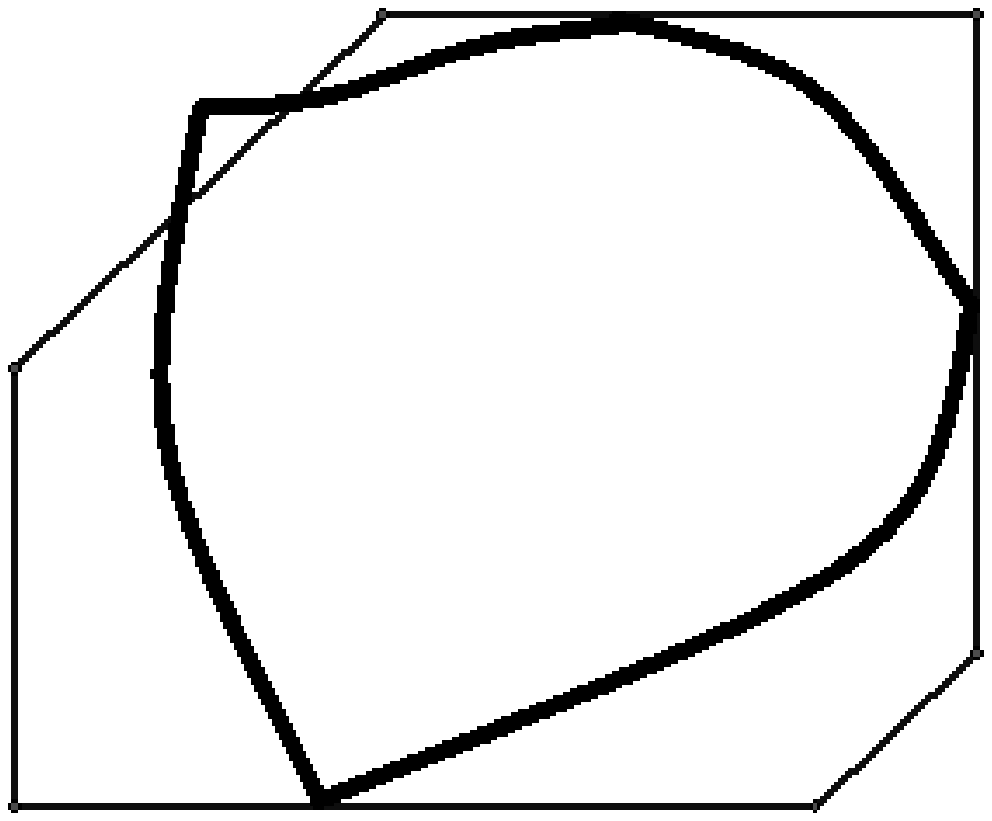}{The situation OUTR7.}{.2}
\figs{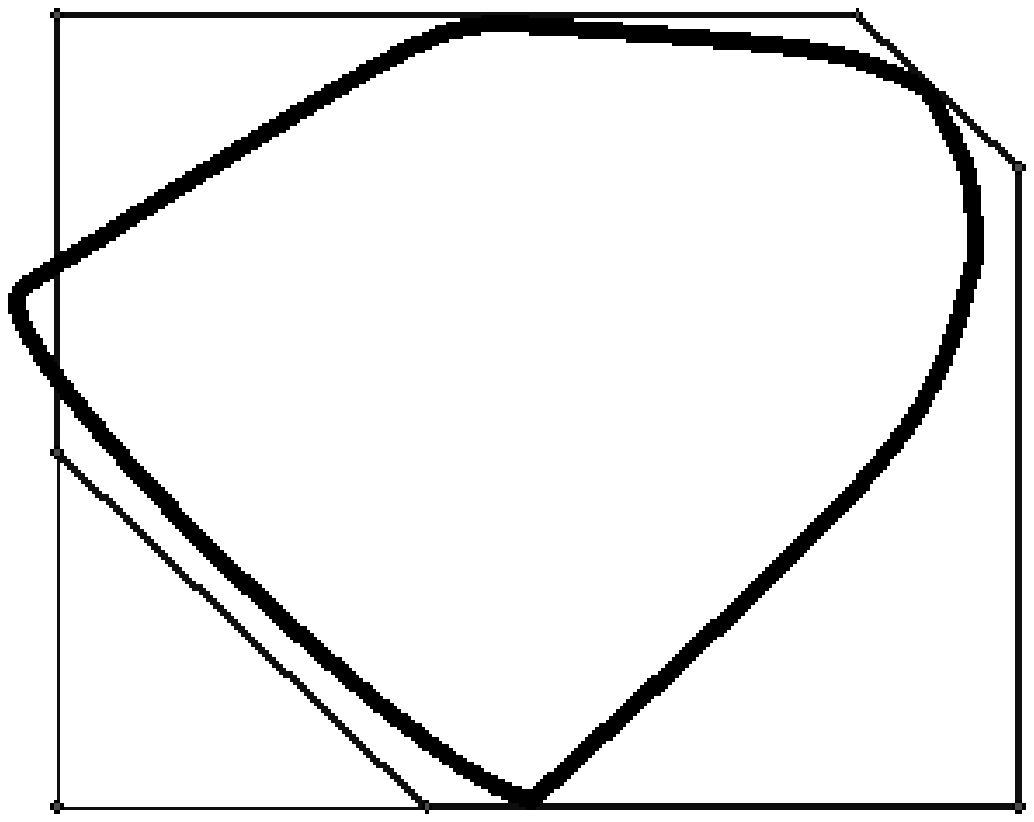}{The situation that $p_5$ is on the border of the small
corner and the arc leaves the box.}{.2}
(see \figref{OUTL5}), and the situation that $p_7$ is in the small corner is called
outL7 (see \figref{outL7}). When the box is moving to the left (the arc is
relatively moving to the right), the situation that $p_7$ is in the big
corner is called OUTR7 (see \figref{OUTR7}), and the situation that $p_5$ is in
the small corner is called outR5 (see \figref{outR5}).
When the box is moving to the situation that $p_5$ is on the border of the
small corner, the arc may leave the box (see \figref{xR=xR5-left}).

\section{The main theorem}

To prove Theorem 1, we start by suppose that a convex closed
unit arc $\gamma$ cannot be covered. We will finally find a contradiction,
mostly by that $\gamma$ is longer than 1.00001, using numerical optimization.

When we translate the box horizontally, consider a big corner and the opposite
small corner, the arc is either passing through the big one or the small
one. When the arc is placed to the left, the points $p_3$ and $p_5$ may visit
the big corners. Similarly for the right justification, the point $p_1$ and
$p_7$ may visit the big corners.
We then divide into cases as follows (see \figref{cases}).

\figs{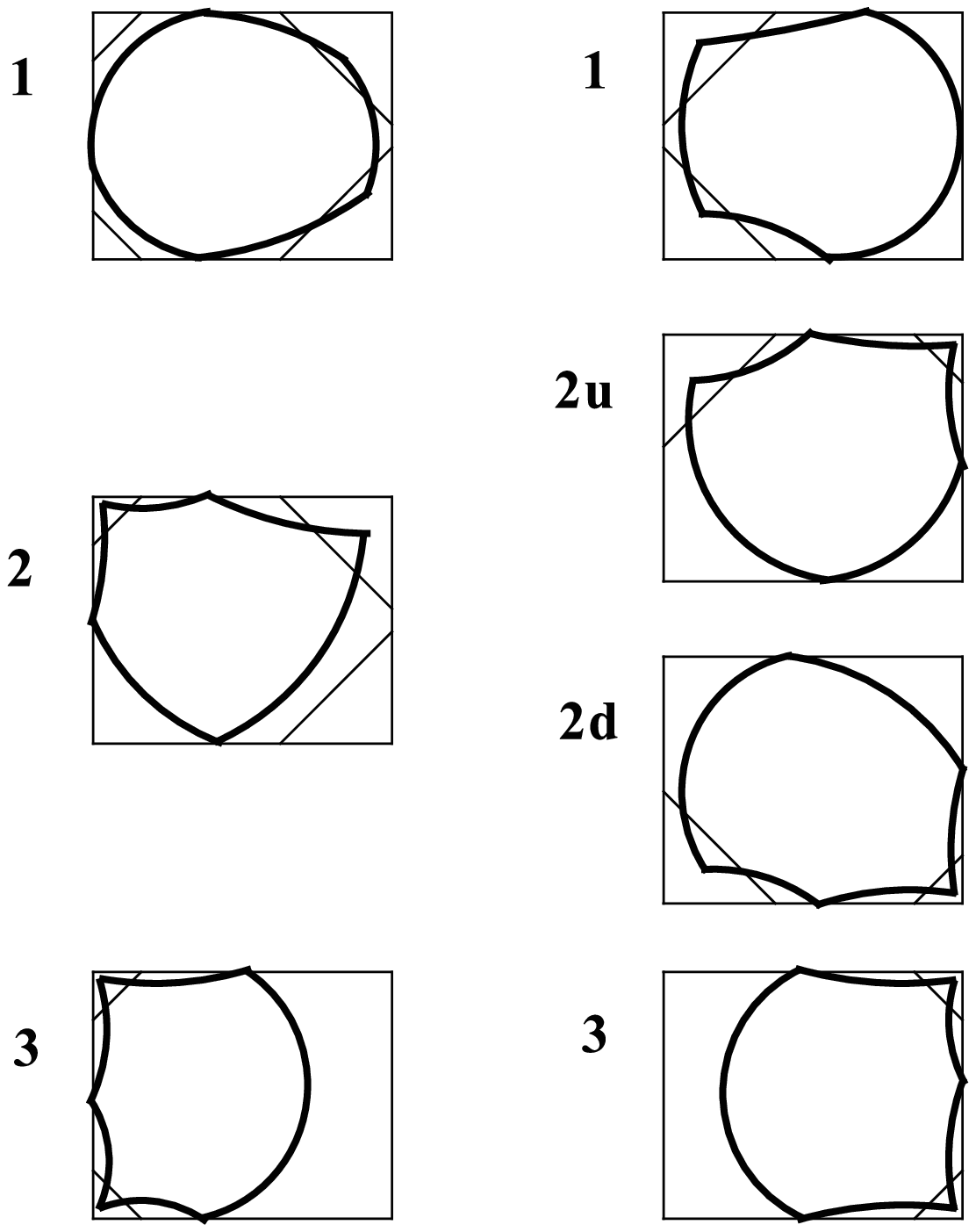}{All cases.}{0.5}

Case 1: OUTL3 and OUTL5.

Case 2: OUTL5 and (not OUTL3, then) outL7.

Case 3: (not OUTL3 nor OUTL5, then) outL1 and outL7.

Now we are going into each big case where we will encounter many subcases.

Case 1: OUTL3 and OUTL5.
Now consider the right justification. Due to the symmetry of case 1, we have
the following subcases.

Subcase 1.1: OUTR7 and OUTR1. From \cite{\fw}, $l(\gamma)\ge l(p_1 p_2 ... p_8)>
1$.

\figs{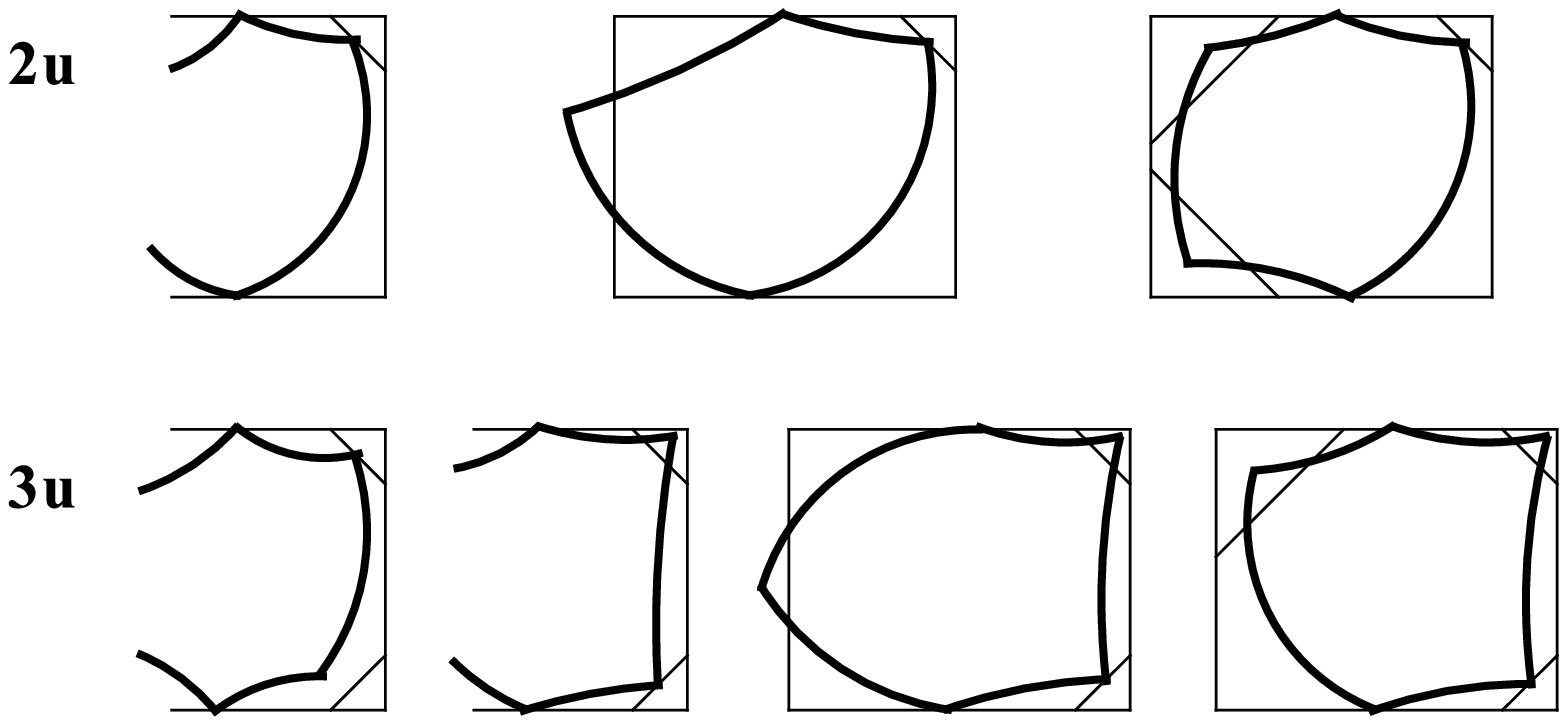}{Translations in subcases ending with .2u and .3u.}{0.5}

Subcase 1.2u: (2up) OUTR7 and (not OUTR1, then) outR5.
Now we translate the box so that $p_5$ is on the border of the small corner
(see \figref{case2u+3u}).
Consequently $x_R=\frac{w-t_2-y_5}{-m}+x_5$, called $\bf x_{R5}$. Then the arc leaves these covers
on the left side. Suppose the arc leave the box on the left side, i.e. $x_8<x_L$,
from numerical work 1.2uL, its length is at least 1.001.
Hence the arc does not leave the box and then it leaves the cover from corners. Thus it is OUT7 and OUT1. Note that we may regard OUTL5 as OUT5 with $x_L=x_8$
and regard OUTR7 as OUT7 with $x_R=x_4$. Here OUT7 and OUT1 are with specific
$x_R$ above. From numerical work 1.2uC, the length is at least 1.02231.

Subcase 1.3: (not OUTR7 nor OUTR1, then) outR3 and outR5.
Now we translate the box so that $p_5$ is on the border of the small corner
and (without of loss of generality) $p_3$ does not visit the small corner.
Suppose the arc does not leave the box, it leaves the cover through corners.
Now it is OUTL7 and OUTL1 with this $x_R$. Now translate the box so that $p_3$ is on the
border of the small corner where $x_R=\frac{t_2-y_3}m+x_3$, called $\bf x_{R3}$. Here $p_5$ visits
the small corner and the arc does not leave the left of the box. Hence it is OUT7 with this $x_R$. From numerical work 1.3C, the length is at least
1.00852.
Now suppose the arc leaves the box when $x_R=x_{R5}$. We translate the box so that $p_3$ is
on the border of the small corner. If the arc does not leave the box, from numerical work 1.3LC, the length is at least 1.00994. If the
arc  leaves the box, from numerical work 1.3LL, the length is at least 1.04008.

Case 2: OUTL5 and (not OUTL3, then) outL7.
We start by translating so that $p_7$ is on the border of the left corner.
Here $x_L=\frac{w-t_2-y_7}m+x_7$, called $\bf x_{L7}$.
Subcases are according to right justification.

Subcase 2.2u: (2up) OUTR7 and (not OUTR1, then) outR5.
Now we translate so that $x_R=x_{R5}$. If both translations keep the arc in
side the boxes, we have OUT3 and OUT5 when $x_L=x_{L7}$ and have OUT1 and
OUT7 when $x_R=x_{R5}$. From numerical work 2C.2uC, the length is at least 1.0093. If the arc leaves the box when $x_R=x_{R5}$, we have OUT1 and OUT7.
(2C.2uL) The length is at least 1.01069.
If the arc leaves the box when $x_L=x_{L7}$ and when $x_R=x_{R5}$, (2R.2uL)
the length is at least 1.03344.

Subcase 2.2d: (2down) OUTR1 and (not OUTR7, then) outR3.
From 2.2d, the length is at least 1.00584.

Subcase 2.3: (not OUTR7 nor OUTR1, then) outR3 and outR5. When we translate
so that $x_R=x_{R5}$. Now we consider whether $p_3$ is in the small corner.

Subsubcase 2.3u: $p_3$ is not in the small corner (see \figref{case2u+3u}). We will translate so that
$x_L=x_{L7}$ and so that $x_R=x_{R5}$. If the arc are in both boxes,
we have OUT3 and OUT5 when $x_L=x_{L7}$. (2C.3uC) The
length is at least 1.00596. If the arc leaves the box when $x_R=x_{R5}$, (2.3uL)
the length is at least 1.00318. If the arc leaves the box when $x_L=x_{L7}$,
(2R.3u) the length is at least 1.00392.

Subsubcase 2.3d: $p_3$ is in the small corner. So we may think that when we
translate so that $x_R=x_{R3}$, we have $p_5$ not in the small corner.
Similarly to the previous subsubcase, from 2C.3dC, 2.3dL and 2R.3d,
we have the length is at least 1.00854, 1.00504 and 1.00392 respectively.

Case 3: (not OUTL3 nor OUTL5, then) outL1 and outL7.
We now have to consider only when the right justification causes (not OUTR7 nor OUTR1, then) outR3 and outR5. We will divide into subcase similar to
the subcase 2.3 but with fewer subcases due to symmetry.

Subcase 3.3u: From 3C.3uC (OUT3 and OUT5 when $x_L$=$x_{L7}$, OUT1 and
OUT7 when $x_R=x_{R5}$) and 3.3uL ($x_8<x_4-l$ when $x_R=x_{R5}$), the length is at least
1.00001 and 1.05382 respectively.

Subcase 3.3d: From 3C.3dC and 3.3dL, the length is at least
1.00001 and 1.05367 respectively.

In every case, the length is greater than 1 (or numerically greater than
1.00001). This is a contradiction. Therefore $\mathfrak{X}$ is a cover for
closed unit arcs. The theorem is proved completely.

\section{Further improvements}
Repeating all the numerical process with threshold smaller than 1.00001 would
yield a
smaller cover.

We tried to make a significant improvement by  subdividing the subsubcases according
to numerical works 3C.3uC and 3C.3dC using the same method in subcase 1.3.
The result is negative.

As the cover has to accommodate the straight back-and-forth segment of length $\frac 12$, it is clearly that there is no way to improve to another smaller
cover by cutting the third corner (clipping 3 corners).

\section{Acknowledgement}
We are grateful to John E. Wetzel for providing the introductory material and the associated references.
We also appreciate the help from Banyat Sroysang. This study is partially supported by the 90th Anniversary of Chu- lalongkorn University Fund (Ratchadaphiseksomphot Endowment Fund).

\section{Appendix: the numerical work}
Each numerical minimization used here is called a \emph{convex programming}
where we minimize a convex function over a convex domain. It is theoritically
confirmed that the minimum may be obtained numerically with high accuracy.

We numerically compute those minimums using Wolfram Mathematica. The Mathematica notebook file can be found at 

\textbf{www.math.sc.chula.ac.th/\textasciitilde wacharin/optimization/closed\%20arcs}

The remaining of this section is the mathematica code and its explation.

Here are the Mathematica notebook for this work. The input cells are in bold-face font while the output cells are in normal font. Note that some graphics outputs are omitted.

\begin{doublespace}
\noindent\(\pmb{d[\text{xys$\_$}]\text{:=}}\\
\pmb{\text{If}\left[\text{Length}[\text{xys}]\geq 2,\surd \left((\text{xys}[[1,1]]-\text{xys}[[2,1]])^2+(\text{xys}[[1,2]]-\text{xys}[[2,2]])^2\right)+d[\text{Rest}[\text{xys}]],\right.}\\
\pmb{0]}\\
\pmb{w=\frac{1}{\pi };}\\
\pmb{l=\sqrt{\frac{1}{4}-\frac{1}{\pi ^2}};}\\
\pmb{s=0.1420171;}\\
\pmb{t=0.1481552;}\\
\pmb{m=\frac{t}{s};}\\
\pmb{\text{s2}=0.0617141;}\\
\pmb{\text{t2}=m \text{s2};}\\
\pmb{\theta =\text{ArcTan}[m];}\\
\pmb{\text{area}=w l-\frac{1}{2}(s t+\text{s2} \text{t2});}\\
\pmb{\text{y2}=0;}\\
\pmb{\text{y6}=w;}\\
\pmb{\text{x8}=0;}\\
\pmb{\text{p1}=\{\text{x1},\text{y1}\};}\\
\pmb{\text{p2}=\{\text{x2},\text{y2}\};}\\
\pmb{\text{p3}=\{\text{x3},\text{y3}\};}\\
\pmb{\text{p4}=\{\text{x4},\text{y4}\};}\\
\pmb{\text{p5}=\{\text{x5},\text{y5}\};}\\
\pmb{\text{p6}=\{\text{x6},\text{y6}\};}\\
\pmb{\text{p7}=\{\text{x7},\text{y7}\};}\\
\pmb{\text{p8}=\{\text{x8},\text{y8}\};}\\
\pmb{\text{arc}=\{\text{p1},\text{p2},\text{p3},\text{p4},\text{p5},\text{p6},\text{p7},\text{p8},\text{p1}\};}\\
\pmb{\text{vars}=\text{Complement}[\text{Flatten}[\text{arc}],\{\text{y2},\text{y6},\text{x8}\}];}\\
\pmb{u[\alpha \_]=\{\text{Cos}[\alpha ],\text{Sin}[\alpha ]\};}\\
\pmb{\text{u1}=u\left[-\frac{\pi }{2}-\theta \right]; \text{u3}=u\left[-\frac{\pi }{2}+\theta \right]; \text{u5}=u\left[\frac{\pi }{2}-\theta \right]; \text{u7}=u\left[\frac{\pi }{2}+\theta \right];}\\
\pmb{\text{dist}=\{l,w-t\}.\text{u5};}\\
\pmb{\text{dist2}=\{l,w-\text{t2}\}.\text{u5};}\\
\pmb{\text{out1R}[\text{xR$\_$},\text{dis$\_$}]=(\text{p1}-\{\text{xR},w\}).\text{u1}>\text{dis};}\\
\pmb{\text{out1L}[\text{xL$\_$},\text{dis$\_$}]=\text{out1R}[\text{xL}+l,\text{dis}];}\\
\pmb{\text{out3L}[\text{xL$\_$},\text{dis$\_$}]=(\text{p3}-\{\text{xL},w\}).\text{u3}>\text{dis};}\\
\pmb{\text{out3R}[\text{xR$\_$},\text{dis$\_$}]=\text{out3L}[\text{xR}-l,\text{dis}];}\\
\pmb{\text{out5L}[\text{xL$\_$},\text{dis$\_$}]=(\text{p5}-\{\text{xL},0\}).\text{u5}>\text{dis};}\\
\pmb{\text{out5R}[\text{xR$\_$},\text{dis$\_$}]=\text{out5L}[\text{xR}-l,\text{dis}];}\\
\pmb{\text{out7R}[\text{xR$\_$},\text{dis$\_$}]=(\text{p7}-\{\text{xR},0\}).\text{u7}>\text{dis};}\\
\pmb{\text{out7L}[\text{xL$\_$},\text{dis$\_$}]=\text{out7R}[\text{xL}+l,\text{dis}];}\\
\pmb{\text{bigout1L}[\text{xL$\_$}]=\text{out1L}[\text{xL},\text{dist}]; \text{bigout1R}[\text{xR$\_$}]=\text{out1R}[\text{xR},\text{dist}];}\\
\pmb{\text{bigout3L}[\text{xL$\_$}]=\text{out3L}[\text{xL},\text{dist}]; \text{bigout3R}[\text{xR$\_$}]=\text{out3R}[\text{xR},\text{dist}];}\\
\pmb{\text{bigout5L}[\text{xL$\_$}]=\text{out5L}[\text{xL},\text{dist}]; \text{bigout5R}[\text{xR$\_$}]=\text{out5R}[\text{xR},\text{dist}];}\\
\pmb{\text{bigout7L}[\text{xL$\_$}]=\text{out7L}[\text{xL},\text{dist}]; \text{bigout7R}[\text{xR$\_$}]=\text{out7R}[\text{xR},\text{dist}];}\\
\pmb{\text{smallout1L}[\text{xL$\_$}]=\text{out1L}[\text{xL},\text{dist2}]; \text{smallout1R}[\text{xR$\_$}]=\text{out1R}[\text{xR},\text{dist2}];}\\
\pmb{\text{smallout3L}[\text{xL$\_$}]=\text{out3L}[\text{xL},\text{dist2}]; \text{smallout3R}[\text{xR$\_$}]=\text{out3R}[\text{xR},\text{dist2}];}\\
\pmb{\text{smallout5L}[\text{xL$\_$}]=\text{out5L}[\text{xL},\text{dist2}]; \text{smallout5R}[\text{xR$\_$}]=\text{out5R}[\text{xR},\text{dist2}];}\\
\pmb{\text{smallout7L}[\text{xL$\_$}]=\text{out7L}[\text{xL},\text{dist2}]; \text{smallout7R}[\text{xR$\_$}]=\text{out7R}[\text{xR},\text{dist2}];}\\
\pmb{\text{big1L}=\text{bigout1L}[0]; \text{big1R}=\text{bigout1R}[\text{x4}];}\\
\pmb{\text{big3L}=\text{bigout3L}[0]; \text{big3R}=\text{bigout3R}[\text{x4}];}\\
\pmb{\text{big5L}=\text{bigout5L}[0]; \text{big5R}=\text{bigout5R}[\text{x4}];}\\
\pmb{\text{big7L}=\text{bigout7L}[0]; \text{big7R}=\text{bigout7R}[\text{x4}];}\\
\pmb{\text{small1L}=\text{smallout1L}[0]; \text{small1R}=\text{smallout1R}[\text{x4}];}\\
\pmb{\text{small3L}=\text{smallout3L}[0]; \text{small3R}=\text{smallout3R}[\text{x4}];}\\
\pmb{\text{small5L}=\text{smallout5L}[0]; \text{small5R}=\text{smallout5R}[\text{x4}];}\\
\pmb{\text{small7L}=\text{smallout7L}[0]; \text{small7R}=\text{smallout7R}[\text{x4}];}\\
\pmb{\text{cond1L}=\text{big3L}\&\&\text{big5L}; }\\
\pmb{\text{cond2L}=\text{big5L}\&\&\text{small7L}; }\\
\pmb{\text{cond3L}=\text{small1L}\&\&\text{small7L};}\\
\pmb{\text{cond1R}=\text{big1R}\&\&\text{big7R}; }\\
\pmb{\text{cond2uR}=\text{big7R}\&\&\text{small5R}; }\\
\pmb{\text{cond2dR}=\text{big1R}\&\&\text{small3R}; }\\
\pmb{\text{cond3R}=\text{small3R}\&\&\text{small5R};}\\
\pmb{\text{xL1}=\frac{\text{t2}-\text{y1}}{-m}+\text{x1};}\\
\pmb{\text{xL7}=\frac{w-\text{t2}-\text{y7}}{m}+\text{x7};}\\
\pmb{\text{xR5}=\frac{w-\text{t2}-\text{y5}}{-m}+\text{x5};}\\
\pmb{\text{xR3}=\frac{\text{t2}-\text{y3}}{m}+\text{x3};}\\
\pmb{\text{big1t3}=\text{bigout1R}[\text{xR3}];}\\
\pmb{\text{big1t5}=\text{bigout1R}[\text{xR5}];}\\
\pmb{\text{big3t1}=\text{bigout3L}[\text{xL1}];}\\
\pmb{\text{big3t7}=\text{bigout3L}[\text{xL7}];}\\
\pmb{\text{big5t1}=\text{bigout5L}[\text{xL1}];}\\
\pmb{\text{big5t7}=\text{bigout5L}[\text{xL7}];}\\
\pmb{\text{big7t3}=\text{bigout7R}[\text{xR3}];}\\
\pmb{\text{big7t5}=\text{bigout7R}[\text{xR5}];}\\
\pmb{\text{leftovert5}=\text{x8}<\text{xR5}-l;}\\
\pmb{\text{leftovert3}=\text{x8}<\text{xR3}-l;}\\
\pmb{\text{rightovert7}=\text{x4}>\text{xL7}+l;}\\
\pmb{\text{rightovert1}=\text{x4}>\text{xL1}+l;}\\
\pmb{\text{bigframeL}[\text{dx$\_$}]=}\\
\pmb{\text{Line}[\{\{s+\text{dx},0\},\{l-s+\text{dx},0\},\{l+\text{dx},t\},\{l+\text{dx},w-t\},\{l+\text{dx}-s,w\},\{s+\text{dx},w\},}\\
\pmb{\{0+\text{dx},w-t\},\{0+\text{dx},t\},\{s+\text{dx},0\}\}];}\\
\pmb{\text{bigframeR}[\text{dx$\_$}]=\text{bigframeL}[\text{dx}-l];}\\
\pmb{\text{smallframeL}[\text{dx$\_$}]=}\\
\pmb{\text{Line}[\{\{\text{s2}+\text{dx},0\},\{l-\text{s2}+\text{dx},0\},\{l+\text{dx},\text{t2}\},\{l+\text{dx},w-\text{t2}\},\{l+\text{dx}-\text{s2},w\},}\\
\pmb{\{\text{s2}+\text{dx},w\},\{0+\text{dx},w-\text{t2}\},\{0+\text{dx},\text{t2}\},\{\text{s2}+\text{dx},0\}\}];}\\
\pmb{\text{smallframeR}[\text{dx$\_$}]=\text{smallframeL}[\text{dx}-l];}\\
\pmb{\text{leftbigframe}=\text{bigframeL}[0]; \text{rightbigframe}=\text{bigframeR}[\text{x4}];}\\
\pmb{\text{leftsmallframe}=\text{smallframeL}[0]; \text{rightsmallframe}=\text{smallframeR}[\text{x4}];}\)
\end{doublespace}

\begin{doublespace}
\noindent\(\pmb{\text{(*} \text{area} \text{of} \text{this} \text{cover} \text{**}\text{**}\text{**}*\text{**}\text{**}\text{**}\text{**}\text{**}\text{**}\text{**}\text{**}\text{**}\text{**}\text{**}\text{**}\text{**}\text{**}\text{**}\text{**}\text{**}\text{**}\text{**}\text{**}\text{**}\text{**}\text{**}\text{**}\text{******)}}\\
\pmb{\text{area}}\)
\end{doublespace}

\begin{doublespace}
\noindent\(0.11023\)
\end{doublespace}

\begin{doublespace}
\noindent\(\pmb{\text{===}\text{===}\text{===}\text{===}\text{===}\text{===}\text{===}\text{===}\text{===}\text{===}\text{===}\text{===}\text{===}\text{===}\text{===}\text{===}\text{===}\text{===}\text{===}\text{==}}\)
\end{doublespace}

\begin{doublespace}
\noindent\(\pmb{\text{(*} 1.2\text{uC} \text{**}\text{**}\text{**}*\text{**}\text{**}\text{**}\text{**}\text{**}\text{**}\text{**}\text{**}\text{**}\text{**}\text{**}\text{**}\text{**}\text{**}\text{**}\text{**}\text{**}\text{**}\text{**}\text{**}\text{**}\text{**}\text{**}\text{**}\text{******)}}\\
\pmb{\text{nm}=\text{NMinimize}[\{d[\text{arc}],\text{cond1L},\text{cond2uR},\text{big1t5}\text{(*}\text{big7t5}\text{*)}\},\text{vars}]}\\
\pmb{\text{Print}[\text{Graphics}[\{\{\text{Red},\text{Line}[\text{arc}]\},\text{leftbigframe}\}]\text{/.}\text{nm}[[2]],}\\
\pmb{\text{Graphics}[\{\{\text{Red},\text{Line}[\text{arc}]\},\text{rightbigframe},\text{rightsmallframe}\}]\text{/.}\text{nm}[[2]],}\\
\pmb{\text{Graphics}[\{\{\text{Red},\text{Line}[\text{arc}]\},\text{bigframeR}[\text{xR5}],\text{smallframeR}[\text{xR5}]\}]\text{/.}\text{nm}[[2]]]}\)
\end{doublespace}

\begin{doublespace}
\noindent\(\{1.02231,\{\text{x1}\to 0.050003,\text{x2}\to 0.190579,\text{x3}\to 0.273266,\text{x4}\to 0.333895,\text{x5}\to 0.309459,\text{x6}\to 0.234306,\text{x7}\to 0.00932146,\text{y1}\to 0.0527391,\text{y3}\to 0.0309776,\text{y4}\to 0.215671,\text{y5}\to 0.290097,\text{y7}\to 0.233807,\text{y8}\to 0.205315\}\}\)
\end{doublespace}

\noindent\(\)

\begin{doublespace}
\noindent\(\pmb{\text{(*} 1.2\text{uL} \text{**}\text{**}\text{**}*\text{**}\text{**}\text{**}\text{**}\text{**}\text{**}\text{**}\text{**}\text{**}\text{**}\text{**}\text{**}\text{**}\text{**}\text{**}\text{**}\text{**}\text{**}\text{**}\text{**}\text{**}\text{**}\text{**}\text{**}\text{******)}}\\
\pmb{\text{nm}=\text{NMinimize}[\{d[\text{arc}],\text{cond1L},\text{cond2uR},\text{leftovert5}\},\text{vars}]}\\
\pmb{\text{Print}[\text{Graphics}[\{\{\text{Red},\text{Line}[\text{arc}]\},\text{leftbigframe}\}]\text{/.}\text{nm}[[2]],}\\
\pmb{\text{Graphics}[\{\{\text{Red},\text{Line}[\text{arc}]\},\text{rightbigframe},\text{rightsmallframe}\}]\text{/.}\text{nm}[[2]],}\\
\pmb{\text{Graphics}[\{\{\text{Red},\text{Line}[\text{arc}]\},\text{bigframeR}[\text{xR5}],\text{smallframeR}[\text{xR5}]\}]\text{/.}\text{nm}[[2]]]}\)
\end{doublespace}

\begin{doublespace}
\noindent\(\{1.001,\{\text{x1}\to 0.159818,\text{x2}\to 0.243572,\text{x3}\to 0.243572,\text{x4}\to 0.325581,\text{x5}\to 0.325579,\text{x6}\to 0.316228,\text{x7}\to -\text{1.2852242374679175$\grave{ }$*${}^{\wedge}$-8},\text{y1}\to 0.0790096,\text{y3}\to -\text{1.713486227141648$\grave{ }$*${}^{\wedge}$-8},\text{y4}\to 0.316522,\text{y5}\to 0.316533,\text{y7}\to 0.232756,\text{y8}\to 0.232756\}\}\)
\end{doublespace}

\noindent\(\)

\begin{doublespace}
\noindent\(\pmb{\text{===}\text{===}\text{===}\text{===}\text{===}\text{===}\text{===}\text{===}\text{===}\text{===}\text{===}\text{===}\text{===}\text{===}\text{===}\text{===}\text{===}\text{===}\text{==}}\)
\end{doublespace}

\begin{doublespace}
\noindent\(\pmb{\text{(*} 1.3C \text{**}\text{**}\text{**}*\text{**}\text{**}\text{**}\text{**}\text{**}\text{**}\text{**}\text{**}\text{**}\text{**}\text{**}\text{**}\text{**}\text{**}\text{**}\text{**}\text{**}\text{**}\text{**}\text{**}\text{**}\text{**}\text{**}\text{**}\text{******)}}\\
\pmb{\text{nm}=\text{NMinimize}[\{d[\text{arc}],\text{cond1L},\text{cond3R},\text{big1t5},\text{big7t5},\text{smallout5R}[\text{xR3}],\text{big7t3}\},\text{vars}]}\\
\pmb{\text{Print}[\text{Graphics}[\{\{\text{Red},\text{Line}[\text{arc}]\},\text{leftbigframe}\}]\text{/.}\text{nm}[[2]],}\\
\pmb{\text{Graphics}[\{\{\text{Red},\text{Line}[\text{arc}]\},\text{rightsmallframe}\}]\text{/.}\text{nm}[[2]],}\\
\pmb{\text{Graphics}[\{\{\text{Red},\text{Line}[\text{arc}]\},\text{bigframeR}[\text{xR5}],\text{smallframeR}[\text{xR5}]\}]\text{/.}\text{nm}[[2]],}\\
\pmb{\text{Graphics}[\{\{\text{Red},\text{Line}[\text{arc}]\},\text{bigframeR}[\text{xR3}],\text{smallframeR}[\text{xR3}]\}]\text{/.}\text{nm}[[2]]]}\)
\end{doublespace}

\begin{doublespace}
\noindent\(\{1.00852,\{\text{x1}\to 0.0188778,\text{x2}\to 0.243563,\text{x3}\to 0.243572,\text{x4}\to 0.243544,\text{x5}\to 0.243573,\text{x6}\to 0.243573,\text{x7}\to 0.0171495,\text{y1}\to 0.0446889,\text{y3}\to -\text{3.931211032843245$\grave{ }$*${}^{\wedge}$-8},\text{y4}\to 0.17721,\text{y5}\to 0.31831,\text{y7}\to 0.271819,\text{y8}\to 0.165333\}\}\)
\end{doublespace}

\noindent\(\)

\begin{doublespace}
\noindent\(\pmb{\text{(*} 1.3\text{LC} \text{**}\text{**}\text{**}*\text{**}\text{**}\text{**}\text{**}\text{**}\text{**}\text{**}\text{**}\text{**}\text{**}\text{**}\text{**}\text{**}\text{**}\text{**}\text{**}\text{**}\text{**}\text{**}\text{**}\text{**}\text{**}\text{**}\text{**}\text{******)}}\\
\pmb{\text{nm}=\text{NMinimize}[\{d[\text{arc}],\text{cond1L},\text{cond3R},\text{leftovert5},\text{smallout5R}[\text{xR3}],\text{big7t3}\},\text{vars}]}\\
\pmb{\text{Print}[\text{Graphics}[\{\{\text{Red},\text{Line}[\text{arc}]\},\text{leftbigframe}\}]\text{/.}\text{nm}[[2]],}\\
\pmb{\text{Graphics}[\{\{\text{Red},\text{Line}[\text{arc}]\},\text{rightsmallframe}\}]\text{/.}\text{nm}[[2]],}\\
\pmb{\text{Graphics}[\{\{\text{Red},\text{Line}[\text{arc}]\},\text{bigframeR}[\text{xR5}],\text{smallframeR}[\text{xR5}]\}]\text{/.}\text{nm}[[2]],}\\
\pmb{\text{Graphics}[\{\{\text{Red},\text{Line}[\text{arc}]\},\text{bigframeR}[\text{xR3}],\text{smallframeR}[\text{xR3}]\}]\text{/.}\text{nm}[[2]]]}\)
\end{doublespace}

\begin{doublespace}
\noindent\(\{1.00994,\{\text{x1}\to 0.0926624,\text{x2}\to 0.252064,\text{x3}\to 0.252064,\text{x4}\to 0.294873,\text{x5}\to 0.323609,\text{x6}\to 0.322395,\text{x7}\to -\text{2.582780802772688$\grave{ }$*${}^{\wedge}$-8},\text{y1}\to 0.154586,\text{y3}\to -\text{2.3090433358166587$\grave{ }$*${}^{\wedge}$-8},\text{y4}\to 0.194448,\text{y5}\to 0.318587,\text{y7}\to 0.245069,\text{y8}\to 0.245069\}\}\)
\end{doublespace}

\noindent\(\)

\begin{doublespace}
\noindent\(\pmb{\text{(*} 1.3\text{LL} \text{**}\text{**}\text{**}*\text{**}\text{**}\text{**}\text{**}\text{**}\text{**}\text{**}\text{**}\text{**}\text{**}\text{**}\text{**}\text{**}\text{**}\text{**}\text{**}\text{**}\text{**}\text{**}\text{**}\text{**}\text{**}\text{**}\text{**}\text{******)}}\\
\pmb{\text{nm}=\text{NMinimize}[\{d[\text{arc}],\text{cond1L},\text{cond3R},\text{leftovert5},\text{smallout5R}[\text{xR3}],\text{leftovert3}\},\text{vars}]}\\
\pmb{\text{Print}[\text{Graphics}[\{\{\text{Red},\text{Line}[\text{arc}]\},\text{leftbigframe}\}]\text{/.}\text{nm}[[2]],}\\
\pmb{\text{Graphics}[\{\{\text{Red},\text{Line}[\text{arc}]\},\text{rightsmallframe}\}]\text{/.}\text{nm}[[2]],}\\
\pmb{\text{Graphics}[\{\{\text{Red},\text{Line}[\text{arc}]\},\text{bigframeR}[\text{xR5}],\text{smallframeR}[\text{xR5}]\}]\text{/.}\text{nm}[[2]],}\\
\pmb{\text{Graphics}[\{\{\text{Red},\text{Line}[\text{arc}]\},\text{bigframeR}[\text{xR3}],\text{smallframeR}[\text{xR3}]\}]\text{/.}\text{nm}[[2]]]}\)
\end{doublespace}

\begin{doublespace}
\noindent\(\{1.04008,\{\text{x1}\to 0.187222,\text{x2}\to 0.323874,\text{x3}\to 0.323875,\text{x4}\to 0.321918,\text{x5}\to 0.323875,\text{x6}\to 0.323875,\text{x7}\to 0.226897,\text{y1}\to 0.0663409,\text{y3}\to -\text{1.1434568252162222$\grave{ }$*${}^{\wedge}$-7},\text{y4}\to 0.17472,\text{y5}\to 0.31831,\text{y7}\to 0.271759,\text{y8}\to 0.159304\}\}\)
\end{doublespace}

\noindent\(\)

\begin{doublespace}
\noindent\(\pmb{\text{===}\text{===}\text{===}\text{===}\text{===}\text{===}\text{===}\text{===}\text{===}\text{===}\text{===}\text{===}\text{===}\text{===}\text{===}\text{===}\text{===}\text{===}\text{===}\text{==}}\)
\end{doublespace}

\begin{doublespace}
\noindent\(\pmb{\text{(*} 2C.2\text{uC} \text{**}\text{**}\text{**}*\text{**}\text{**}\text{**}\text{**}\text{**}\text{**}\text{**}\text{**}\text{**}\text{**}\text{**}\text{**}\text{**}\text{**}\text{**}\text{**}\text{**}\text{**}\text{**}\text{**}\text{**}\text{**}\text{**}\text{**}\text{******)}}\\
\pmb{\text{nm}=\text{NMinimize}[\{d[\text{arc}],\text{cond2L},\text{cond2uR},\text{big3t7}\text{(*}\text{big5t7}\text{*)},\text{big1t5}\text{(*}\text{big7t5}\text{*)}\},\text{vars}]}\\
\pmb{\text{Print}[\text{Graphics}[\{\{\text{Red},\text{Line}[\text{arc}]\},\text{leftbigframe},\text{leftsmallframe}\}]\text{/.}\text{nm}[[2]],}\\
\pmb{\text{Graphics}[\{\{\text{Red},\text{Line}[\text{arc}]\},\text{rightbigframe},\text{rightsmallframe}\}]\text{/.}\text{nm}[[2]],}\\
\pmb{\text{Graphics}[\{\{\text{Red},\text{Line}[\text{arc}]\},\text{bigframeL}[\text{xL7}],\text{smallframeL}[\text{xL7}]\}]\text{/.}\text{nm}[[2]],}\\
\pmb{\text{Graphics}[\{\{\text{Red},\text{Line}[\text{arc}]\},\text{bigframeR}[\text{xR5}],\text{smallframeR}[\text{xR5}]\}]\text{/.}\text{nm}[[2]]]}\)
\end{doublespace}

\begin{doublespace}
\noindent\(\{1.0093,\{\text{x1}\to 0.0473402,\text{x2}\to 0.122243,\text{x3}\to 0.198457,\text{x4}\to 0.245602,\text{x5}\to 0.245602,\text{x6}\to 0.102025,\text{x7}\to -\text{4.596229860767697$\grave{ }$*${}^{\wedge}$-8},\text{y1}\to 0.0149951,\text{y3}\to 0.0151983,\text{y4}\to 0.316179,\text{y5}\to 0.316192,\text{y7}\to 0.316192,\text{y8}\to 0.316191\}\}\)
\end{doublespace}

\noindent\(\)

\begin{doublespace}
\noindent\(\pmb{\text{(*} 2C.2\text{uL} \text{**}\text{**}\text{**}*\text{**}\text{**}\text{**}\text{**}\text{**}\text{**}\text{**}\text{**}\text{**}\text{**}\text{**}\text{**}\text{**}\text{**}\text{**}\text{**}\text{**}\text{**}\text{**}\text{**}\text{**}\text{**}\text{**}\text{**}\text{******)}}\\
\pmb{\text{nm}=\text{NMinimize}[\{d[\text{arc}],\text{cond2L},\text{cond2uR},\text{big3t7}\text{(*}\text{big5t7}\text{*)},\text{leftovert5}\},\text{vars}]}\\
\pmb{\text{Print}[\text{Graphics}[\{\{\text{Red},\text{Line}[\text{arc}]\},\text{leftbigframe},\text{leftsmallframe}\}]\text{/.}\text{nm}[[2]],}\\
\pmb{\text{Graphics}[\{\{\text{Red},\text{Line}[\text{arc}]\},\text{rightbigframe},\text{rightsmallframe}\}]\text{/.}\text{nm}[[2]],}\\
\pmb{\text{Graphics}[\{\{\text{Red},\text{Line}[\text{arc}]\},\text{bigframeL}[\text{xL7}],\text{smallframeL}[\text{xL7}]\}]\text{/.}\text{nm}[[2]],}\\
\pmb{\text{Graphics}[\{\{\text{Red},\text{Line}[\text{arc}]\},\text{bigframeR}[\text{xR5}],\text{smallframeR}[\text{xR5}]\}]\text{/.}\text{nm}[[2]]]}\)
\end{doublespace}

\begin{doublespace}
\noindent\(\{1.01069,\{\text{x1}\to 0.0410888,\text{x2}\to 0.242837,\text{x3}\to 0.242837,\text{x4}\to 0.312801,\text{x5}\to 0.315407,\text{x6}\to 0.266864,\text{x7}\to -\text{3.634189465709122$\grave{ }$*${}^{\wedge}$-8},\text{y1}\to 0.2109,\text{y3}\to -\text{3.138234567372148$\grave{ }$*${}^{\wedge}$-8},\text{y4}\to 0.315837,\text{y5}\to 0.327144,\text{y7}\to 0.254695,\text{y8}\to 0.254695\}\}\)
\end{doublespace}

\noindent\(\)

\begin{doublespace}
\noindent\(\pmb{\text{(*} 2R.2\text{uL} \text{**}\text{**}\text{**}*\text{**}\text{**}\text{**}\text{**}\text{**}\text{**}\text{**}\text{**}\text{**}\text{**}\text{**}\text{**}\text{**}\text{**}\text{**}\text{**}\text{**}\text{**}\text{**}\text{**}\text{**}\text{**}\text{**}\text{**}\text{******)}}\\
\pmb{\text{nm}=\text{NMinimize}[\{d[\text{arc}],\text{cond2L},\text{cond2uR},\text{rightovert7},\text{leftovert5}\},\text{vars}]}\\
\pmb{\text{Print}[\text{Graphics}[\{\{\text{Red},\text{Line}[\text{arc}]\},\text{leftbigframe},\text{leftsmallframe}\}]\text{/.}\text{nm}[[2]],}\\
\pmb{\text{Graphics}[\{\{\text{Red},\text{Line}[\text{arc}]\},\text{rightbigframe},\text{rightsmallframe}\}]\text{/.}\text{nm}[[2]],}\\
\pmb{\text{Graphics}[\{\{\text{Red},\text{Line}[\text{arc}]\},\text{bigframeL}[\text{xL7}],\text{smallframeL}[\text{xL7}]\}]\text{/.}\text{nm}[[2]],}\\
\pmb{\text{Graphics}[\{\{\text{Red},\text{Line}[\text{arc}]\},\text{bigframeR}[\text{xR5}],\text{smallframeR}[\text{xR5}]\}]\text{/.}\text{nm}[[2]]]}\)
\end{doublespace}

\begin{doublespace}
\noindent\(\{1.03344,\{\text{x1}\to 0.0351126,\text{x2}\to 0.173417,\text{x3}\to 0.242828,\text{x4}\to 0.34657,\text{x5}\to 0.34657,\text{x6}\to 0.159403,\text{x7}\to -\text{5.617573323650674$\grave{ }$*${}^{\wedge}$-8},\text{y1}\to 0.234658,\text{y3}\to 0.118697,\text{y4}\to 0.294634,\text{y5}\to 0.294634,\text{y7}\to 0.294634,\text{y8}\to 0.294634\}\}\)
\end{doublespace}

\noindent\(\)

\begin{doublespace}
\noindent\(\pmb{\text{===}\text{===}\text{===}\text{===}\text{===}\text{===}\text{===}\text{===}\text{===}\text{===}\text{===}\text{===}\text{===}\text{===}\text{===}\text{===}\text{===}\text{===}\text{===}\text{==}}\)
\end{doublespace}

\begin{doublespace}
\noindent\(\pmb{\text{(*} 2.2d \text{**}\text{**}\text{**}*\text{**}\text{**}\text{**}\text{**}\text{**}\text{**}\text{**}\text{**}\text{**}\text{**}\text{**}\text{**}\text{**}\text{**}\text{**}\text{**}\text{**}\text{**}\text{**}\text{**}\text{**}\text{**}\text{**}\text{**}\text{******)}}\\
\pmb{\text{nm}=\text{NMinimize}[\{d[\text{arc}],\text{cond2L},\text{cond2dR}\},\text{vars}]}\\
\pmb{\text{Print}[\text{Graphics}[\{\{\text{Red},\text{Line}[\text{arc}]\},\text{leftbigframe},\text{leftsmallframe}\}]\text{/.}\text{nm}[[2]],}\\
\pmb{\text{Graphics}[\{\{\text{Red},\text{Line}[\text{arc}]\},\text{rightbigframe},\text{rightsmallframe}\}]\text{/.}\text{nm}[[2]]]}\)
\end{doublespace}

\begin{doublespace}
\noindent\(\{1.00584,\{\text{x1}\to -\text{2.3136767549496994$\grave{ }$*${}^{\wedge}$-6},\text{x2}\to -\text{1.7910053485531854$\grave{ }$*${}^{\wedge}$-6},\text{x3}\to 0.21252,\text{x4}\to 0.243569,\text{x5}\to 0.243572,\text{x6}\to 0.243122,\text{x7}\to 0.0313227,\text{y1}\to -\text{7.002475819513364$\grave{ }$*${}^{\wedge}$-8},\text{y3}\to 0.0319897,\text{y4}\to 0.318271,\text{y5}\to 0.31831,\text{y7}\to 0.286605,\text{y8}\to 0.000087863\}\}\)
\end{doublespace}

\noindent\(\)

\begin{doublespace}
\noindent\(\pmb{\text{===}\text{===}\text{===}\text{===}\text{===}\text{===}\text{===}\text{===}\text{===}\text{===}\text{===}\text{===}\text{===}\text{===}\text{===}\text{===}\text{===}\text{===}\text{===}\text{==}}\)
\end{doublespace}

\begin{doublespace}
\noindent\(\pmb{\text{(*} 2C.3\text{uC} \text{**}\text{**}\text{**}*\text{**}\text{**}\text{**}\text{**}\text{**}\text{**}\text{**}\text{**}\text{**}\text{**}\text{**}\text{**}\text{**}\text{**}\text{**}\text{**}\text{**}\text{**}\text{**}\text{**}\text{**}\text{**}\text{**}\text{**}\text{******)}}\\
\pmb{\text{nm}=\text{NMinimize}[\{d[\text{arc}],\text{cond2L},\text{cond3R},\text{big3t7}\text{(*}\text{big5t7}\text{*)},\text{big1t5},\text{big7t5}\},\text{vars}]}\\
\pmb{\text{Print}[\text{Graphics}[\{\{\text{Red},\text{Line}[\text{arc}]\},\text{leftbigframe},\text{leftsmallframe}\}]\text{/.}\text{nm}[[2]],}\\
\pmb{\text{Graphics}[\{\{\text{Red},\text{Line}[\text{arc}]\},\text{rightbigframe},\text{rightsmallframe}\}]\text{/.}\text{nm}[[2]],}\\
\pmb{\text{Graphics}[\{\{\text{Red},\text{Line}[\text{arc}]\},\text{bigframeL}[\text{xL7}],\text{smallframeL}[\text{xL7}]\}]\text{/.}\text{nm}[[2]],}\\
\pmb{\text{Graphics}[\{\{\text{Red},\text{Line}[\text{arc}]\},\text{bigframeR}[\text{xR5}],\text{smallframeR}[\text{xR5}]\}]\text{/.}\text{nm}[[2]]]}\)
\end{doublespace}

\begin{doublespace}
\noindent\(\{1.00596,\{\text{x1}\to 0.0321129,\text{x2}\to 0.21358,\text{x3}\to 0.213724,\text{x4}\to 0.217391,\text{x5}\to 0.243572,\text{x6}\to 0.243571,\text{x7}\to 0.0000588819,\text{y1}\to 0.0308805,\text{y3}\to 0.0000181782,\text{y4}\to 0.0404414,\text{y5}\to 0.31831,\text{y7}\to 0.285145,\text{y8}\to 0.284457\}\}\)
\end{doublespace}

\noindent\(\)

\begin{doublespace}
\noindent\(\pmb{\text{(*} 2.3\text{uL} \text{**}\text{**}\text{**}*\text{**}\text{**}\text{**}\text{**}\text{**}\text{**}\text{**}\text{**}\text{**}\text{**}\text{**}\text{**}\text{**}\text{**}\text{**}\text{**}\text{**}\text{**}\text{**}\text{**}\text{**}\text{**}\text{**}\text{**}\text{******)}}\\
\pmb{\text{nm}=\text{NMinimize}[\{d[\text{arc}],\text{cond2L},\text{cond3R},\text{leftovert5},\text{x5}\leq \text{x4}\},\text{vars}]}\\
\pmb{\text{Print}[\text{Graphics}[\{\{\text{Red},\text{Line}[\text{arc}]\},\text{leftbigframe},\text{leftsmallframe}\}]\text{/.}\text{nm}[[2]],}\\
\pmb{\text{Graphics}[\{\{\text{Red},\text{Line}[\text{arc}]\},\text{rightbigframe},\text{rightsmallframe}\}]\text{/.}\text{nm}[[2]],}\\
\pmb{\text{Graphics}[\{\{\text{Red},\text{Line}[\text{arc}]\},\text{bigframeR}[\text{xR5}],\text{smallframeR}[\text{xR5}]\}]\text{/.}\text{nm}[[2]]]}\)
\end{doublespace}

\begin{doublespace}
\noindent\(\{1.00318,\{\text{x1}\to 0.138809,\text{x2}\to 0.197367,\text{x3}\to 0.197367,\text{x4}\to 0.259081,\text{x5}\to 0.259081,\text{x6}\to 0.126285,\text{x7}\to -\text{2.7060825018296567$\grave{ }$*${}^{\wedge}$-8},\text{y1}\to 0.0743581,\text{y3}\to \text{6.517086758888646$\grave{ }$*${}^{\wedge}$-9},\text{y4}\to 0.385583,\text{y5}\to 0.385904,\text{y7}\to 0.253928,\text{y8}\to 0.253928\}\}\)
\end{doublespace}

\noindent\(\)

\begin{doublespace}
\noindent\(\pmb{\text{(*} 2R.3u \text{**}\text{**}\text{**}*\text{**}\text{**}\text{**}\text{**}\text{**}\text{**}\text{**}\text{**}\text{**}\text{**}\text{**}\text{**}\text{**}\text{**}\text{**}\text{**}\text{**}\text{**}\text{**}\text{**}\text{**}\text{**}\text{**}\text{**}\text{******)}}\\
\pmb{\text{nm}=\text{NMinimize}[\{d[\text{arc}],\text{cond2L},\text{cond3R},\text{rightovert7},\text{y7}\leq \text{y6}\},\text{vars}]}\\
\pmb{\text{Print}[\text{Graphics}[\{\{\text{Red},\text{Line}[\text{arc}]\},\text{leftbigframe},\text{leftsmallframe}\}]\text{/.}\text{nm}[[2]],}\\
\pmb{\text{Graphics}[\{\{\text{Red},\text{Line}[\text{arc}]\},\text{rightbigframe},\text{rightsmallframe}\}]\text{/.}\text{nm}[[2]],}\\
\pmb{\text{Graphics}[\{\{\text{Red},\text{Line}[\text{arc}]\},\text{bigframeL}[\text{xL7}],\text{smallframeL}[\text{xL7}]\}]\text{/.}\text{nm}[[2]]]}\)
\end{doublespace}

\begin{doublespace}
\noindent\(\{1.00392,\{\text{x1}\to 0.178549,\text{x2}\to 0.253095,\text{x3}\to 0.253095,\text{x4}\to 0.314809,\text{x5}\to 0.314801,\text{x6}\to -0.00830716,\text{x7}\to -0.00906586,\text{y1}\to 0.0904813,\text{y3}\to -\text{1.1828094866898152$\grave{ }$*${}^{\wedge}$-8},\text{y4}\to 0.253814,\text{y5}\to 0.253937,\text{y7}\to 0.31831,\text{y8}\to 0.307586\}\}\)
\end{doublespace}

\noindent\(\)

\begin{doublespace}
\noindent\(\pmb{\text{===}\text{===}\text{===}\text{===}\text{===}\text{===}\text{===}\text{===}\text{===}\text{===}\text{===}\text{===}\text{===}\text{===}\text{===}\text{===}\text{===}\text{===}\text{===}\text{==}}\)
\end{doublespace}

\begin{doublespace}
\noindent\(\pmb{\text{(*} 2C.3\text{dC} \text{**}\text{**}\text{**}*\text{**}\text{**}\text{**}\text{**}\text{**}\text{**}\text{**}\text{**}\text{**}\text{**}\text{**}\text{**}\text{**}\text{**}\text{**}\text{**}\text{**}\text{**}\text{**}\text{**}\text{**}\text{**}\text{**}\text{**}\text{******)}}\\
\pmb{\text{nm}=\text{NMinimize}[\{d[\text{arc}],\text{cond2L},\text{cond3R},\text{big3t7}\text{(*}\text{big5t7}\text{*)},\text{big1t3},\text{big7t3}\},\text{vars}]}\\
\pmb{\text{Print}[\text{Graphics}[\{\{\text{Red},\text{Line}[\text{arc}]\},\text{leftbigframe},\text{leftsmallframe}\}]\text{/.}\text{nm}[[2]],}\\
\pmb{\text{Graphics}[\{\{\text{Red},\text{Line}[\text{arc}]\},\text{rightbigframe},\text{rightsmallframe}\}]\text{/.}\text{nm}[[2]],}\\
\pmb{\text{Graphics}[\{\{\text{Red},\text{Line}[\text{arc}]\},\text{bigframeL}[\text{xL7}],\text{smallframeL}[\text{xL7}]\}]\text{/.}\text{nm}[[2]],}\\
\pmb{\text{Graphics}[\{\{\text{Red},\text{Line}[\text{arc}]\},\text{bigframeR}[\text{xR3}],\text{smallframeR}[\text{xR3}]\}]\text{/.}\text{nm}[[2]]]}\)
\end{doublespace}

\begin{doublespace}
\noindent\(\{1.00854,\{\text{x1}\to 0.0196674,\text{x2}\to 0.243574,\text{x3}\to 0.243574,\text{x4}\to 0.24364,\text{x5}\to 0.243572,\text{x6}\to 0.243564,\text{x7}\to 0.0155361,\text{y1}\to 0.0438665,\text{y3}\to -\text{1.0641765711192708$\grave{ }$*${}^{\wedge}$-8},\text{y4}\to 0.0810013,\text{y5}\to 0.31831,\text{y7}\to 0.270136,\text{y8}\to 0.175562\}\}\)
\end{doublespace}

\noindent\(\)

\begin{doublespace}
\noindent\(\pmb{\text{(*} 2.3\text{dL} \text{**}\text{**}\text{**}*\text{**}\text{**}\text{**}\text{**}\text{**}\text{**}\text{**}\text{**}\text{**}\text{**}\text{**}\text{**}\text{**}\text{**}\text{**}\text{**}\text{**}\text{**}\text{**}\text{**}\text{**}\text{**}\text{**}\text{**}\text{******)}}\\
\pmb{\text{nm}=\text{NMinimize}[\{d[\text{arc}],\text{cond2L},\text{cond3R},\text{leftovert3},\text{x3}\leq \text{x4}\},\text{vars}]}\\
\pmb{\text{Print}[\text{Graphics}[\{\{\text{Red},\text{Line}[\text{arc}]\},\text{leftbigframe},\text{leftsmallframe}\}]\text{/.}\text{nm}[[2]],}\\
\pmb{\text{Graphics}[\{\{\text{Red},\text{Line}[\text{arc}]\},\text{rightbigframe},\text{rightsmallframe}\}]\text{/.}\text{nm}[[2]],}\\
\pmb{\text{Graphics}[\{\{\text{Red},\text{Line}[\text{arc}]\},\text{bigframeR}[\text{xR3}],\text{smallframeR}[\text{xR3}]\}]\text{/.}\text{nm}[[2]]]}\)
\end{doublespace}

\begin{doublespace}
\noindent\(\{1.00504,\{\text{x1}\to 0.0662786,\text{x2}\to 0.294492,\text{x3}\to 0.310332,\text{x4}\to 0.310332,\text{x5}\to 0.248714,\text{x6}\to 0.248098,\text{x7}\to -\text{5.7530756332999174$\grave{ }$*${}^{\wedge}$-8},\text{y1}\to 0.198461,\text{y3}\to -0.0141285,\text{y4}\to -0.013705,\text{y5}\to 0.318209,\text{y7}\to 0.253928,\text{y8}\to 0.253928\}\}\)
\end{doublespace}

\noindent\(\)

\begin{doublespace}
\noindent\(\pmb{\text{(*} 2R.3d \text{**}\text{**}\text{**}*\text{**}\text{**}\text{**}\text{**}\text{**}\text{**}\text{**}\text{**}\text{**}\text{**}\text{**}\text{**}\text{**}\text{**}\text{**}\text{**}\text{**}\text{**}\text{**}\text{**}\text{**}\text{**}\text{**}\text{**}\text{******)}}\\
\pmb{\text{nm}=\text{NMinimize}[\{d[\text{arc}],\text{cond2L},\text{cond3R},\text{rightovert7},\text{y7}\leq \text{y6}\},\text{vars}]}\\
\pmb{\text{Print}[\text{Graphics}[\{\{\text{Red},\text{Line}[\text{arc}]\},\text{leftbigframe},\text{leftsmallframe}\}]\text{/.}\text{nm}[[2]],}\\
\pmb{\text{Graphics}[\{\{\text{Red},\text{Line}[\text{arc}]\},\text{rightbigframe},\text{rightsmallframe}\}]\text{/.}\text{nm}[[2]],}\\
\pmb{\text{Graphics}[\{\{\text{Red},\text{Line}[\text{arc}]\},\text{bigframeL}[\text{xL7}],\text{smallframeL}[\text{xL7}]\}]\text{/.}\text{nm}[[2]]]}\)
\end{doublespace}

\begin{doublespace}
\noindent\(\{1.00392,\{\text{x1}\to 0.178549,\text{x2}\to 0.253095,\text{x3}\to 0.253095,\text{x4}\to 0.314809,\text{x5}\to 0.314801,\text{x6}\to -0.00830716,\text{x7}\to -0.00906586,\text{y1}\to 0.0904813,\text{y3}\to -\text{1.1828094866898152$\grave{ }$*${}^{\wedge}$-8},\text{y4}\to 0.253814,\text{y5}\to 0.253937,\text{y7}\to 0.31831,\text{y8}\to 0.307586\}\}\)
\end{doublespace}

\noindent\(\)

\begin{doublespace}
\noindent\(\pmb{\text{===}\text{===}\text{===}\text{===}\text{===}\text{===}\text{===}\text{===}\text{===}\text{===}\text{===}\text{===}\text{===}\text{===}\text{===}\text{===}\text{===}\text{===}\text{===}\text{==}}\)
\end{doublespace}

\begin{doublespace}
\noindent\(\pmb{\text{(*} 3C.3\text{uC} \text{**}\text{**}\text{**}*\text{**}\text{**}\text{**}\text{**}\text{**}\text{**}\text{**}\text{**}\text{**}\text{**}\text{**}\text{**}\text{**}\text{**}\text{**}\text{**}\text{**}\text{**}\text{**}\text{**}\text{**}\text{**}\text{**}\text{**}\text{******)}}\\
\pmb{\text{nm}=\text{NMinimize}[\{d[\text{arc}],\text{cond3L},\text{cond3R},\text{big3t7},\text{big5t7},\text{big1t5},\text{big7t5}\},\text{vars}]}\\
\pmb{\text{Print}[\text{Graphics}[\{\{\text{Red},\text{Line}[\text{arc}]\},\text{leftbigframe},\text{leftsmallframe}\}]\text{/.}\text{nm}[[2]],}\\
\pmb{\text{Graphics}[\{\{\text{Red},\text{Line}[\text{arc}]\},\text{rightbigframe},\text{rightsmallframe}\}]\text{/.}\text{nm}[[2]],}\\
\pmb{\text{Graphics}[\{\{\text{Red},\text{Line}[\text{arc}]\},\text{bigframeL}[\text{xL1}],\text{smallframeL}[\text{xL1}]\}]\text{/.}\text{nm}[[2]],}\\
\pmb{\text{Graphics}[\{\{\text{Red},\text{Line}[\text{arc}]\},\text{bigframeR}[\text{xR5}],\text{smallframeR}[\text{xR5}]\}]\text{/.}\text{nm}[[2]]]}\)
\end{doublespace}

\begin{doublespace}
\noindent\(\{1.00001,\{\text{x1}\to -0.000154244,\text{x2}\to 0.067712,\text{x3}\to 0.189222,\text{x4}\to 0.189243,\text{x5}\to 0.189379,\text{x6}\to 0.0629569,\text{x7}\to -0.0000795552,\text{y1}\to 0.00278395,\text{y3}\to 0.00510414,\text{y4}\to 0.0891912,\text{y5}\to 0.313087,\text{y7}\to 0.315649,\text{y8}\to 0.114689\}\}\)
\end{doublespace}

\noindent\(\)

\begin{doublespace}
\noindent\(\pmb{\text{{``}1.00001{''}}}\)
\end{doublespace}

\begin{doublespace}
\noindent\(\pmb{\text{(*} 3.3\text{uL} \text{**}\text{**}\text{**}*\text{**}\text{**}\text{**}\text{**}\text{**}\text{**}\text{**}\text{**}\text{**}\text{**}\text{**}\text{**}\text{**}\text{**}\text{**}\text{**}\text{**}\text{**}\text{**}\text{**}\text{**}\text{**}\text{**}\text{**}\text{******)}}\\
\pmb{\text{nm}=\text{NMinimize}[\{d[\text{arc}],\text{cond3L},\text{cond3R},\text{leftovert5},\text{x5}\leq \text{x4}\},\text{vars}]}\\
\pmb{\text{Print}[\text{Graphics}[\{\{\text{Red},\text{Line}[\text{arc}]\},\text{leftbigframe},\text{leftsmallframe}\}]\text{/.}\text{nm}[[2]],}\\
\pmb{\text{Graphics}[\{\{\text{Red},\text{Line}[\text{arc}]\},\text{rightbigframe},\text{rightsmallframe}\}]\text{/.}\text{nm}[[2]],}\\
\pmb{\text{Graphics}[\{\{\text{Red},\text{Line}[\text{arc}]\},\text{bigframeR}[\text{xR5}],\text{smallframeR}[\text{xR5}]\}]\text{/.}\text{nm}[[2]]]}\)
\end{doublespace}

\begin{doublespace}
\noindent\(\{1.05382,\{\text{x1}\to 0.0245756,\text{x2}\to 0.142258,\text{x3}\to 0.142265,\text{x4}\to 0.203978,\text{x5}\to 0.203978,\text{x6}\to 0.0617142,\text{x7}\to 0.061714,\text{y1}\to 0.0387436,\text{y3}\to \text{1.329925959155903$\grave{ }$*${}^{\wedge}$-6},\text{y4}\to 0.41927,\text{y5}\to 0.443389,\text{y7}\to 0.31831,\text{y8}\to 0.120877\}\}\)
\end{doublespace}

\noindent\(\)

\begin{doublespace}
\noindent\(\pmb{\text{===}\text{===}\text{===}\text{===}\text{===}\text{===}\text{===}\text{===}\text{===}\text{===}\text{===}\text{===}\text{===}\text{===}\text{===}\text{===}\text{===}\text{===}\text{===}\text{==}}\)
\end{doublespace}

\begin{doublespace}
\noindent\(\pmb{\text{(*} 3C.3\text{dC} \text{**}\text{**}\text{**}*\text{**}\text{**}\text{**}\text{**}\text{**}\text{**}\text{**}\text{**}\text{**}\text{**}\text{**}\text{**}\text{**}\text{**}\text{**}\text{**}\text{**}\text{**}\text{**}\text{**}\text{**}\text{**}\text{**}\text{**}\text{******)}}\\
\pmb{\text{nm}=\text{NMinimize}[\{d[\text{arc}],\text{cond3L},\text{cond3R},\text{big3t7},\text{big5t7},\text{big1t3},\text{big7t3}\},\text{vars}]}\\
\pmb{\text{Print}[\text{Graphics}[\{\{\text{Red},\text{Line}[\text{arc}]\},\text{leftbigframe},\text{leftsmallframe}\}]\text{/.}\text{nm}[[2]],}\\
\pmb{\text{Graphics}[\{\{\text{Red},\text{Line}[\text{arc}]\},\text{rightbigframe},\text{rightsmallframe}\}]\text{/.}\text{nm}[[2]],}\\
\pmb{\text{Graphics}[\{\{\text{Red},\text{Line}[\text{arc}]\},\text{bigframeL}[\text{xL1}],\text{smallframeL}[\text{xL1}]\}]\text{/.}\text{nm}[[2]],}\\
\pmb{\text{Graphics}[\{\{\text{Red},\text{Line}[\text{arc}]\},\text{bigframeR}[\text{xR3}],\text{smallframeR}[\text{xR3}]\}]\text{/.}\text{nm}[[2]]]}\)
\end{doublespace}

\begin{doublespace}
\noindent\(\{1.00001,\{\text{x1}\to -0.0000710035,\text{x2}\to 0.18117,\text{x3}\to 0.18934,\text{x4}\to 0.189314,\text{x5}\to 0.189344,\text{x6}\to 0.000754577,\text{x7}\to -0.000152621,\text{y1}\to 0.00754623,\text{y3}\to 0.000333404,\text{y4}\to 0.163008,\text{y5}\to 0.310378,\text{y7}\to 0.318273,\text{y8}\to 0.101918\}\}\)
\end{doublespace}

\noindent\(\)

\begin{doublespace}
\noindent\(\pmb{\text{{``}1.00001{''}}}\)
\end{doublespace}

\begin{doublespace}
\noindent\(\pmb{\text{(*} 3.3\text{dL} \text{**}\text{**}\text{**}*\text{**}\text{**}\text{**}\text{**}\text{**}\text{**}\text{**}\text{**}\text{**}\text{**}\text{**}\text{**}\text{**}\text{**}\text{**}\text{**}\text{**}\text{**}\text{**}\text{**}\text{**}\text{**}\text{**}\text{**}\text{******)}}\\
\pmb{\text{nm}=\text{NMinimize}[\{d[\text{arc}],\text{cond3L},\text{cond3R},\text{leftovert3},\text{x3}\leq \text{x4}\},\text{vars}]}\\
\pmb{\text{Print}[\text{Graphics}[\{\{\text{Red},\text{Line}[\text{arc}]\},\text{leftbigframe},\text{leftsmallframe}\}]\text{/.}\text{nm}[[2]],}\\
\pmb{\text{Graphics}[\{\{\text{Red},\text{Line}[\text{arc}]\},\text{rightbigframe},\text{rightsmallframe}\}]\text{/.}\text{nm}[[2]],}\\
\pmb{\text{Graphics}[\{\{\text{Red},\text{Line}[\text{arc}]\},\text{bigframeR}[\text{xR3}],\text{smallframeR}[\text{xR3}]\}]\text{/.}\text{nm}[[2]]]}\)
\end{doublespace}

\begin{doublespace}
\noindent\(\{1.05367,\{\text{x1}\to 0.0617141,\text{x2}\to 0.0617142,\text{x3}\to 0.20732,\text{x4}\to 0.20732,\text{x5}\to 0.14561,\text{x6}\to 0.145571,\text{x7}\to 0.0221245,\text{y1}\to -\text{1.0022566385933313$\grave{ }$*${}^{\wedge}$-8},\text{y3}\to -0.121592,\text{y4}\to -0.108602,\text{y5}\to 0.318306,\text{y7}\to 0.277009,\text{y8}\to 0.200147\}\}\)
\end{doublespace}

\textbf{The following is the explanation of the mathematica code.}

\setlength{\parindent}{0em}
\setlength{\parskip}{1em}
d is the length function of any polygonal arc

w and l are the dimension of the cover by Schaer and Wetzel (fig. 1)

s and t are the side lengths by Furedi and Wetzel (fig. 1)

m is the slope t/s

s2 is the side length in this work (fig. 2)

t2 is the side length in this work with the slope t2/s2=m

$\theta$ is the angle from slope m to X-axis

area is the area of this new cover

p1,p2,...,p8 are points with coordinates (xi,yi) (fig. 3) we set y2=0, y6=w, x8=0

arc is the polygonal closed arc

u[$\alpha$] is the unit vector of angle $\alpha$ (to X-axis)

Let s1,s2,...,s8 be the support lines of points p1,p2,...,p8 respectively. Their slopes are -m,0,m,$\infty$,-m,0,m,$\infty$ respectively.

u1,u3,u5,u7 are unit vectors pointing outward and perpendicular to s1,s3,s5,s7 respectively (fig. 3)

dist is the distant from (l,w) to the lower-left side of the new cover in fig. 3 (we can see this as the length of vector (l,w-t) in direction of u5)

dist2 is the distance from (0,0) to the upper-right side of the new cover in fig. 3 (we can see this as the length of vector (l,w-t2) in direction of -u1

out1R[xR,dis] is the condition that the length of vector p1-(xR,w) in direction of u1 is greater than dis (will be use to describe that p1 is going to escape the lower-left corner when the arc is justified right at x=xR)

out1L[xL,dis] is the condition that the length of vector p1-(xL+l,w) in direction of u1 is greater than dis (will be use to describe that p1 is going to escape the lower-left corner when the arc is justified left at x=xL)

out3L[xL,dis] is the condition that the length of vector p3-(xL,w) in direction of u3 is greater than dis

out3R[xR,dis] is the condition that the length of vector p3-(xR-l,w) in direction of u3 is greater than dis

out5L[xL,dis] is the condition that the length of vector p5-(xL,0) in direction of u5 is greater than dis

out5R[xR,dis] is the condition that the length of vector p5-(xR-l,0) in direction of u5 is greater than dis

out7R[xR,dis] is the condition that the length of vector p7-(xR,0) in direction of u7 is greater than dis

out7L[xL,dis] is the condition that the length of vector p7-(xL+l,0) in direction of u7 is greater than dis

bigout1L[xL] is the condition out1L[xL,dist] (equivalent to that p1 escapes the lower-left big corner when the arc is justified left at x=xL)

bigout1R[xR] is the condition out1R[xR,dist] (equivalent to that p1 escapes the lower-left big corner when the arc is justified right at x=xR)

Similarly for other bigoutkL/bigoutkR conditions.

smallout1L[xL] is the condition out1L[xL,dist2] (equivalent to that p1 escapes the lower-left small corner when the arc is justified left at x=xL)

smallout1R[xR] is equivalent to that p1 escapes the lower-left small corner when the arc is justified right at x=xR

Similarly for smalloutkL/smalloutkR conditions.

big1L is the condition that p1 escapes the big corner when justified left at x=0

big1R is the condition that p1 escapes the big corner when justified right at x=x4

Similarly for bigkL/bigkR conditions.

small1L is the condition that p1 escapes the small corner when justified left at x=0

small1R is the condition that p1 escapes the small corner when justified right at x=x4

Similarly, for smallkL/smallkR conditions.

Now wer mention the main conditions used in the article.

cond1L is big3L and big5L (upper left of fig. 9) \\
cond2L is big5L and small7L (middle left of fig. 9) \\
cond3L is small1L and small7L (lower left of fig. 9) \\
cond1R is big1R and big7R (upper right of fig. 9) \\ 
cond2uR is big7R and small5R (2u of fig. 9) \\ 
cond2dR is big1R and small3R (2d of fig. 9) \\ 
cond3R is small3R and small5R (lower right of fig. 9)

xL1 = (t2 - y1)/-m + x1  is the minimum x of the arc when justified to the lower-left small corner

xL7 = (w - t2 - y7)/m + x7  is the minimum x of the arc when justified to the upper-left small corner

xR5 = (w - t2 - y5)/-m + x5  is the maximum x of the arc when justified to the upper-right small corner

xR3 = (t2 - y3)/m + x3  is the maximum x of the arc when justified to the lower-right small corner

big1t3 is bigout1R[xR3] \\
big1t5 is bigout1R[xR5] \\
big3t1 is bigout3L[xL1] \\
big3t7 is bigout3L[xL7] \\
big5t1 is bigout5L[xL1] \\
big5t7 is bigout5L[xL7] \\
big7t3 is bigout7R[xR3] \\
big7t5 is bigout7R[xR5]

leftovert5 is that x8 $<$ xR5 - l \\
leftovert3 is that x8 $<$ xR3 - l \\
rightovert7 is that x4 $>$ xL7 + l \\
rightovert1 is that x4 $>$ xL1 + l

bigframeL[dx] is the drawing of 8-sided frame with big corners removed at x=dx on the left

bigframeR[dx] is the drawing of 8-sided frame with big corners removed at x=dx on the right

smallframk is the one with small corners removed

leftbigframe is bigframeL[0], justified left at x=0

rightbigframe is bigframeR[x4], justified right at x=x4

leftsmallframe is smallframeL[0]

rightsmallframe is smallframeR[x4]

\textbf{From the article, we translate conditions as follows}

For additional conditions in fig. 10,

In the upper row, .2u: xR=xR5, .2uL: x8<xL, .2uC: OUT1 and OUT7 \\
In the bottom row, .3u: xR=xR3, .3uL: x8<xL, .3uC: OUT7

Hence we have a similar additional conditions as follows.

.2d: xR=xR3, .2dL: x8<xL, .2dC: OUT1 and OUT7 \\
.3d: xR=xR5, .3dL: x8<xL, .3dC: OUT1

Thus the conditions are as follows

1.: cond1L \\
2.: cond2L \\
2C.: cond2L, big3t7, big5t7 \\
2R.: cond2L, rightovert7 \\
3.: cond3L \\
3C. : cond3L, big3t7, big5t7 \\
.2uC : cond2uR, big1t5, big7t5 \\
.2uL : cond2uR, leftovert5 \\
.2d : cond2dR \\
.3C : cond3R, big1t5, big7t5, smallout5R[xR3], big7t3 \\
.3LC : cond3R, leftovert5, smallout5R[xR3], big7t3 \\
.3LL : cond3R, leftovert5, smallout5R[xR3], leftovert3 \\
.3u : cond3R, rightovert7 \\
.3uC : cond3R, big7t3 \\
.3uL : cond3R, leftovert3 \\
.3d : cond3R, rightovert7 \\
.3dC : cond3R, big1t5 \\
.3dL : cond3R, leftovert5 

Now we get into each case and subcase with some redundant conditions omitted in ().

1.2uC: cond1L, cond2uR, big1t5 (big7t5) \\
1.2uL: cond1L, cond2uR, leftovert5 \\
1.3C: cond1L, cond3R, big1t5, big7t5, smallout5R[xR3], big7t3 \\
1.3LC: cond1L, cond3R, leftovert5, smallout5R[xR3], big7t3 \\
1.3LL: cond1L, cond3R, leftovert5, smallout5R[xR3], leftovert3 \\
2C.2uC: cond2L, cond2uR, big3t7 (big5t7), big1t5 (big7t5) \\
2C.2uL: cond2L, cond2uR, big3t7 (big5t7), leftovert5 \\
2R.2uL: cond2L, cond2uR, rightovert7, leftovert5 \\
2.2d: cond2L, cond2dR \\
2C.3uC: cond2L, cond3R, big3t7 (big5t7), big1t5, big7t5 \\
2.3uL: cond2L, cond3R, leftovert5, x5 $\le$ x4 \\
2R.3u: cond2L, cond3R, rightovert7, y7 $\le$ y6 \\
2C.3dC: cond2L, cond3R, big3t7 (big5t7), big1t3, big7t3 \\
2.3dL: cond2L, cond3R, leftovert3, x3 $\le$ x4 \\
2R.3d: cond2L, cond3R, rightovert7, y7 $\le$ y6 \\
3C.3uC: cond3L, cond3R, big3t7, big5t7, big1t5, big7t5 \\
3.3uL: cond3L, cond3R, leftovert5, x5 $\le$ x4 \\
3C.3dC: cond3L, cond3R, big3t7, big5t7, big1t3, big7t3 \\
3.3dL: cond3L, cond3R, leftovert3, x3 $\le$ x4 

Note that some necessary condtions on xi and yi are used. When we omit some conditions, the minimum is even smaller (but still greater than 1.00001). The optimization is of the type called convex programming.

In each case, the output is the minimum length of the critical arc and drawing of the arc when justified to sides of the new cover. The minimum values are as follows (by Mathematica 7.0 and for later versions the result are just a bit different).

1.2uC: 1.02231 \\
1.2uL:  1.001\\
1.3C: 1.00852 \\
1.3LC: 1.00994 \\
1.3LL: 1.04008 \\
2C.2uC: 1.0093 \\
2C.2uL: 1.01069 \\
2R.2uL: 1.03344 \\
2.2d: 1.00584 \\
2C.3uC: 1.00596 \\
2.3uL: 1.00318 \\
2R.3u: 1.00392 \\
2C.3dC: 1.00854 \\
2.3dL: 1.00504 \\
2R.3d: 1.00392 \\
3C.3uC: 1.00001 (also >1.00001 by Mathematica version 11)\\
3.3uL: 1.05382 \\
3C.3dC: 1.00001 (also >1.00001 by Mathematica version 11)\\
3.3dL: 1.05367 \\



\end{document}